\documentclass[aps,prl,superscriptaddress,twocolumn,preprintnumbers,floatfix]{revtex4-2}
\usepackage{color}
\usepackage{dsfont}
\usepackage{mathtools}
\usepackage{amsthm}
\usepackage{amsfonts,amssymb,amsmath}
\usepackage[hidelinks]{hyperref}
\usepackage{physics}
\usepackage{comment}

\usepackage{graphicx}
 \newtheorem{theorem}{Theorem}
  
    \newtheorem{corollary}{Corollary}

  \definecolor{mypurple}{rgb}{0.49,0.18,0.56}

\begin{document}

\title{Complete Hilbert-Space Ergodicity in \\ Quantum Dynamics of Generalized Fibonacci Drives}
\author{Sa\'ul Pilatowsky-Cameo}
\affiliation{Center for Theoretical Physics, Massachusetts Institute of Technology, Cambridge, MA 02139, USA}
\author{Ceren B.~Dag}
\affiliation{ITAMP, Harvard-Smithsonian Center for Astrophysics, Cambridge, MA 02138, USA}
\affiliation{Department of Physics, Harvard University, 17 Oxford Street Cambridge, MA 02138, USA}
\author{Wen Wei Ho}
\email{wenweiho@nus.edu.sg}
\affiliation{Department of Physics, National University of Singapore, Singapore 117542}
\affiliation{Centre for Quantum Technologies, National University of Singapore, 3 Science Drive 2, Singapore 117543}
\author{Soonwon Choi}
\email{soonwon@mit.edu}
\affiliation{Center for Theoretical Physics, Massachusetts Institute of Technology, Cambridge, MA 02139, USA}

\preprint{MIT-CTP/5569}

\begin{abstract}
Ergodicity of quantum dynamics is often defined through statistical properties of energy eigenstates, as exemplified by Berry's conjecture in single-particle quantum chaos and the eigenstate thermalization hypothesis in many-body settings. 
In this work, we investigate whether quantum systems can exhibit a stronger form of ergodicity, wherein any time-evolved state uniformly visits the entire Hilbert space over time. 
We call such a phenomenon {\it complete Hilbert-space ergodicity} (CHSE), which is more akin to the intuitive notion of ergodicity as an inherently dynamical concept.
CHSE cannot hold for time-independent or even time-periodic Hamiltonian dynamics, owing to the existence of (quasi)energy eigenstates which precludes exploration of the full Hilbert space. 
However, we find that there exists a family of aperiodic, yet deterministic drives with minimal symbolic complexity ---  generated by the Fibonacci word and its generalizations --- for which CHSE can be proven to occur. 
Our results provide a basis for understanding thermalization in general time-dependent quantum systems.
\end{abstract}

\maketitle
\newcommand{\E}[0]{\mathop{{}\mathbb{E}}}

One of the cornerstones of statistical physics is the concept of ergodicity: it provides a mechanism by which a generic physical system achieves equilibrium, allowing for a simple statistical description of otherwise complex dynamics.
First put forth by Boltzmann for classical systems, ergodicity refers to their property wherein 
all available states are explored over time, irrespective of the initial configuration~\cite{BookBoltzmann1896}.
For quantum systems, such a dynamical formulation of ergodicity is, however, incongruous with the existence of stationary solutions to Schr\"odinger's equation, such as energy eigenstates.
Instead, quantum ergodicity is often defined through the statistical properties of stationary states. 
In single-particle chaotic systems, Berry's conjecture states that highly-excited energy eigenstates are locally indistinguishable from a superposition of random plane waves~\cite{Berry1977}; and in interacting many-body systems, the eigenstate thermalization hypothesis (ETH) states that most eigenstates behave like random vectors at the level of local observables~\cite{Deutsch1991,Srednicki1994,Rigol2008,Deutsch2018,Kazuya2020}.

This formulation of quantum ergodicity, centered around stationary states of dynamics, is not fully satisfactory.
For quantum systems governed by general time-dependent Hamiltonians, stationary states are typically not well-defined~\cite{Jauslin1991,Blekher1992}.
In fact, energy eigenstates are guaranteed to exist only for systems with  time-independent, or time-periodic Hamiltonians~\cite{Shirley1965}. 
This immediately begs the following questions: is it possible to define a notion of quantum ergodicity that is suitable for a closed quantum system with general time dependence? Do such systems equilibrate or thermalize, and if so, what is the underlying mechanism?

\begin{figure}[t]
\centering\includegraphics[width=1\columnwidth]{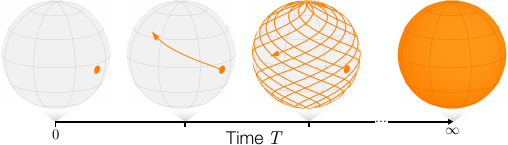}
    \caption{Main idea of complete Hilbert-space ergodicity, illustrated for the case of a single qubit: {\it any} initial state (orange disk) explores the Bloch sphere uniformly over time.}
    \label{fig:fig1}
\end{figure}

In this work, we report progress towards understanding quantum ergodicity in general time-dependent Hamiltonian dynamics, leveraging concepts from quantum information theory.
We propose a stronger form of quantum ergodicity that better captures Boltzmann's original notion of ergodicity as a dynamical property, without reference to stationary states.
Furthermore, we provide a class of simple physical systems where such ergodicity holds.

We define {\it complete Hilbert-space ergodicity} (CHSE) as a property of quantum dynamics for which
the evolution of {\it any} initial state uniformly explores every point of its Hilbert space over time, as depicted in Fig.~\ref{fig:fig1}. 
Equivalently, it is the condition that a time-evolved wavefunction is statistically indistinguishable from random vectors sampled uniformly from the Hilbert space. 
Intuitively, one expects dynamics under a random sequence of unitaries to exhibit CHSE, since any wavefunction will almost surely cover the entire Hilbert space over time. This is analogous to saying Brownian motion is ergodic~\cite{Jian2022}. A more interesting question is whether the same level of ergodicity can be achieved in systems with simple time dependence. Intriguingly, we find the answer in the affirmative.

We prove that CHSE holds in a large family of aperiodic yet deterministic drives, obtained by sequentially applying one of two fixed unitaries in a certain order.
This order is determined by simple concatenation rules that generate the well-known Fibonacci word and its variants. 
Importantly, the complexity of the time-dependence is the minimal possible for CHSE, in a quantifiable way, in contrast to the maximally complex random drives.

Our work has several physical and conceptual implications.
For a many-body system displaying CHSE,  it follows that a local subsystem necessarily achieves thermalization to infinite temperatures for almost-all late times~\cite{Popescu2006}. In fact, it also implies a stronger form of universality that has been recently introduced, called deep thermalization \cite{Claeys2022,Wilming2022,Ho2022,Ippoliti2022,Choi2023,Cotler2023,Ippoliti2023,Lucas2023,Shrotriya2023, NoteSuppmat}. Furthermore, we speculate that quantum systems exhibiting CHSE   constitute arenas where the growth of circuit complexity may be rigorously investigated, a subject of much interest in the quantum information community~\cite{Susskind2016, Haferkamp2022}.

{\it Complete Hilbert-space ergodicity.---}%
Consider a quantum state $|\psi(0)\rangle$ in a finite $d$-dimensional Hilbert space, undergoing dynamics $|\psi(0)\rangle \mapsto |\psi(t)\rangle$ by some Hamiltonian $H(t)$ or sequence of unitary gates. Our formalism below is agnostic of the underlying structure or basis of the Hilbert space. It could be, for example, a fermionic or bosonic Fock space with a fixed number of particles, or a tensor product of many qubits or spins.
We would like to inquire if the state `explores' its ambient space equally likely in time. Mathematically, this can be captured by asking whether the infinite-time average of any integrable function $f$ of the time-evolved state equals its uniform average over states in the Hilbert space, 
 \begin{equation}
        { \lim_{T\to \infty}\frac{1}{T}\int_0^T \dd{t}f(\dyad{\psi(t)}) }={\int
    \dd{\phi} 
        f(\dyad{\phi})},
        \label{eqn:f_avg}
\end{equation}
where $\dd\phi$ is the unique unitarily-invariant measure on the Hilbert space, induced by the Haar measure on the unitary group $\operatorname{SU}(d)$. 
If Eq.~\eqref{eqn:f_avg} is true for any initial state $|\psi(0)\rangle$, then we deem the quantum dynamics as {\it completely Hilbert-space ergodic}. 
Equation~\eqref{eqn:f_avg} is reminiscent of the celebrated Birkhoff's Ergodic Theorem, which is generally applied to classical dynamical systems \cite{Cornfeld1982}.
We are proposing to adopt Eq.~\eqref{eqn:f_avg} as a definition of ergodicity for closed quantum dynamics. 

We systematically characterize CHSE by considering polynomial functions of degree $k$.
For example, the quadratic function $f_O(|\psi(t)\rangle \langle \psi(t)|)=\langle \psi(t)|O|\psi(t)\rangle^2$ gives us information about the temporal fluctuation of an observable $O$.
Equality of Eq.~\eqref{eqn:f_avg} for any polynomial of degree $k$ amounts to probing that the $k$-th moments of the temporal and spatial ensembles, $\rho^{(k)}_T \coloneqq  \frac{1}{T} \int_0^T dt |\psi(t)\rangle \langle \psi(t)|^{\otimes k}$ and $\rho^{(k)}_\text{Haar} \coloneqq  \int d\phi |\phi\rangle \langle \phi|^{\otimes k}$, respectively, match at late times. This agreement can be quantitatively captured by the vanishing of the trace distance
\begin{align}
\Delta^{(k)}_\infty\coloneqq \tfrac{1}{2}\norm*{\rho_\infty^{(k)}-\rho_{\mathrm{Haar}}^{(k)}}_1,
\end{align}
where $\rho^{(k)}_\infty = \lim_{T \to \infty} \rho^{(k)}_T$.
CHSE is achieved if for any initial state $\Delta_\infty^{(k)}=0$  for all $k$ \footnote{This follows because any integrable function on a compact space can be arbitrarily well-approximated by polynomials, a consequence of the Stone-Weierstrass theorem.}.

It can be readily shown that quantum dynamics in a sufficiently large Hilbert space admitting (quasi-)energy eigenstates cannot achieve CHSE.
This is intuitive: the conservation of population on stationary states prevents a complete exploration of the Hilbert space. In the supplemental material (SM) \cite{NoteSuppmat}, we show that for evolution under a time-independent Hamiltonian,
the trace distance for $k \geq 2$ can be always lower bounded as
    \begin{equation}
    \label{eq:boundtmindep}
        \Delta^{(k)}_\infty\geq B(d)\coloneqq  (d+1)^{-1}-(2d(d+1))^{-1/2}>0,
    \end{equation}
irrespective of initial state.
Thus, CHSE can only occur for time-dependent dynamics. A similar obstruction can be shown in time-periodic dynamics~\cite{Pilatowsky2024}  or if the Hamiltonian becomes time-independent after a finite time.

A simple example of a (discrete) time-dependent system that does exhibit CHSE is as follows: let $A_0$ and $A_1$ be two typical unitaries on a $d$-dimensional space, and generate dynamics by randomly applying $A_0$ or $A_1$ with equal probability at each time step. With probability 1, such an evolution exhibits CHSE.
However, this is unsurprising and can be intuitively understood using results from complexity and quantum information theory.
First, it is well known that a pair of unitaries drawn independently and uniformly from the special unitary group $\operatorname{SU}(d)$ is almost surely \emph{quantum computationally universal} \cite{Lloyd1995}, i.e., any other unitary in $\operatorname{SU}(d)$ can be asymptotically reached by some long product of the pair. 
Second, an infinitely long random binary sequence almost surely exhibits a \emph{symbolic complexity} (explained below) which is maximal.
This implies that any possible product of $A_0$ and $A_1$ eventually appears in the sequence.
Combined together, a random sequence of two typical unitaries over long timescales is therefore asymptotically equivalent to a sequence of Haar-random unitaries, which trivially achieves CHSE~\cite{Emerson2005,NoteSuppmat}.

    \begin{figure}[t]
    \centering
\includegraphics[width=1\columnwidth]{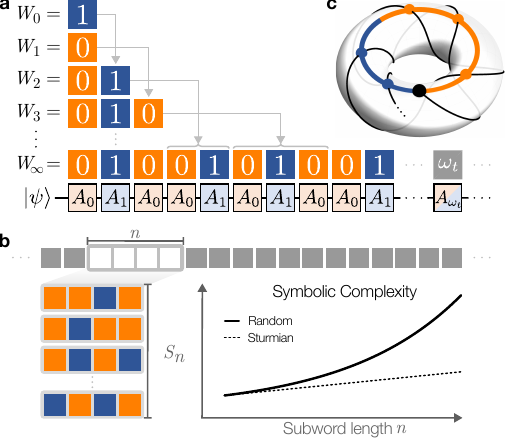}
    \caption{\textbf{a} Sequence of Fibonacci words  constructed by the concatenation rule $W_{j+1}=W_{j} W_{j-1}$, for order $m=1$. The Fibonacci drive applies a pair of unitaries $A_0,A_1$ in the sequence prescribed by $W_\infty$.  \textbf{b} Symbolic complexity $S_n$ counts the number of subwords of length $n$ that appear in an infinite word. It is linear in $n$ for Sturmian words such as $W_\infty$ and exponential in $n$ for random words. \textbf{c} Quasiperiodicity of the Fibonacci drive. Consider a point moving in a straight line (black) which wraps around the torus. Each time the line crosses the orange region, the unitary $A_0$ is applied, and when crosses the blue region, $A_1$ is applied \cite{NoteSuppmat}.}
    \label{fig:fig2}
\end{figure}

{\it Generalized Fibonacci drive.---}%
Here, we introduce a family of nonperiodic, yet deterministic  drives, where despite being generated by simple rules,   CHSE can be shown to hold. 
 Fix a natural number $m$ and two initial words $W_{0}=1$ and $W_{1}=0$. Then, we define a sequence of words by recursively concatenating shorter ones 
 $W_{j+1}=(W_{j})^m W_{j-1}$, where   multiplication should be understood as concatenation of words.
 This process leads to a well defined, infinite word $W_\infty$.
The case $m=1$ is the well-known Fibonacci word \cite{oeisFibb},
$W_\infty=0100101001001\dots,$ (see Fig.~\ref{fig:fig2}a), while the case $m=2$   generates the so-called Pell word  $W_\infty=0010010001001\dots$ \cite{oeisPell}. 
We  thus refer to $W_\infty$ as the  Fibonacci word of order $m$. 
All such sequences are examples of so-called Sturmian words~\cite{BookPytheasFogg2006,DeLuca1997,NoteSuppmat}.

Sturmian words represent the simplest possible aperiodic sequences.
Specifically, they are characterized by having the minimal possible symbolic complexity $S_n$. 
Symbolic complexity measures the number of distinct contiguous subwords of length $n$ that appear in an infinite word \cite{Ferenczi1999}~(Fig.~\ref{fig:fig2}b).
In the case of a random binary word, $S_n=2^n$ (exponential), which is maximal.
In contrast, Sturmian words have complexity  $S_n=n+1$ (linear), which is minimal for any aperiodic word; a word with complexity $S_n < n+1$ can be shown to be eventually repeating~\cite{BookPytheasFogg2006}.

We use the Fibonacci words to generate discrete-time quantum dynamics on a $d$-level system. Given two unitaries $A_0,A_1\in \operatorname{SU}(d)$, we apply $A_{\omega_t}$ at time $t \in \mathbb{N}$,  where $\omega_t$ is the $t$-th symbol in $W_\infty$.
Explicitly, the time evolution operator is given by $U(0)=\mathds{1}$ and
\newcommand{\floor}[1]{\left\lfloor #1 \right\rfloor}
\begin{align}
  U(t)=A_{\omega_{t}}\cdots\,A_{\omega_3}A_{\omega_2}A_{\omega_1}\label{eq:evoloperatordefinition}
\end{align}
for $t\geq 1$, so that a time-evolved state is $|\psi(t)\rangle=U(t)|\psi(0)\rangle$.
We call $U(t)$ the generalized Fibonacci drive of order $m$. 
This drive may be understood as arising from a time-quasiperiodic Hamiltonian \cite{Ho1983,Luck1988,Verdeny2016,Crowley2019,Dominic2020}, as its time dependence can be shown to factor through a $2$-dimensional torus (Fig.~\ref{fig:fig2}c)~\cite{NoteSuppmat}. The Fibonacci drive of order $1$ has gained recent attention in the quantum-dynamics community~\cite{Dumitrescu2018,Mori2021,Long2022,Zhao2022} and has been experimentally realized \cite{Dumitrescu2022}.
  
 {\it CHSE in the Fibonacci drives.---} %
 Because the generalized Fibonacci drives are defined in discrete time, we modify the condition for CHSE in Eq.~\eqref{eqn:f_avg} and  temporal ensembles accordingly, changing the integral over continuous time to a sum over discrete time, e.g.,~  $\rho_T^{(k)}=\frac{1}{T}\sum_{t=0}^{T-1}|\psi(t)\rangle\langle\psi(t)|^{\otimes{k}}.$
Our aim is to show $\rho_T^{(k)}$ converges to $\rho_\textrm{Haar}^{(k)}$ for large $T$, independent of initial state.
 
 To show this, we note that $\rho^{(k)}_T$ can be obtained by passing the (replicated)  initial state  $\rho^{(k)}_0=|\psi(0)\rangle\langle\psi(0)|^{\otimes{k}}$ through a time-averaging quantum channel, 
 \begin{align}\label{eq:TAC}\mathcal{N}^{(k)}_T[\,\cdot\,]\coloneqq\frac{1}{T}\sum_{t=0}^{T-1}U(t)^{\otimes{k}}(\,\cdot\,)U(t)^{\dagger\otimes{k}},
    \end{align}
such that $\rho^{(k)}_\infty=\mathcal{N}^{(k)}_\infty[\rho^{(k)}_0]$, where $\mathcal{N}^{(k)}_\infty=\lim_{T\to\infty} \mathcal{N}_T^{(k)}$ is the infinite-time averaging channel.
Here, a subtle technical point has to be noted: in general, it is not guaranteed that the limit $\mathcal{N}^{(k)}_\infty$ should exist  \footnote{One can come up with contrived examples of quantum dynamics where $\mathcal{N}^{(k)}_\infty$ does not converge.}. 
Using Birkhoff's Ergodic Theorem, we proved that it exists for a class of quasiperiodic Hamiltonians related to the Fibonacci drives by an initial phase shift on the torus~\cite{NoteSuppmat}, but we could not show  this for the Fibonacci drives themselves.
Extensive numerics performed below, however, suggest that it does.
We proceed with the assumption that the $\mathcal{N}^{(k)}_\infty$ is well defined in this class of models. The question is then what this limit is. We have: 

\begin{theorem} 
For almost all pairs of unitaries $A_0, A_1 \in \operatorname{SU}(d)$, 
$\mathcal{N}^{(k)}_\infty$ of the Fibonacci drive of order $m$ (assuming it exists), satisfies
\label{th:01}   
     \begin{align}\forall k\in\mathbb{N}:\quad\mathcal{N}^{(k)}_\infty=\mathcal{N}_{\mathrm{Haar}}^{(k)},
\end{align}
where $\mathcal{N}_{\mathrm{Haar}}^{(k)}[\,\cdot\,]=\int\dd{U}U^{\otimes k}(\,\cdot\,)U^{\dagger \otimes k}$ is the $k$-fold Haar-averaging channel.
 \end{theorem}

This theorem means that for almost all $A_0$ and $A_1$ (barring exceptional scenarios such as $A_0=A_1=\mathds{1}$), time-averaging under quantum dynamics  of the Fibonacci drives is equivalent to a randomization  under uniformly distributed unitaries:  the evolution operators 
$\{U(t)\}_{t\in\mathbb{N}}$ 
are statistically indistinguishable from Haar-random unitaries. 

Since $\mathcal{N}^{(k)}_\text{Haar}[\rho^{(k)}_0]=\rho^{(k)}_\text{Haar}$ for any initial state, this establishes our main result:
\begin{corollary} 
\label{corollary:CHSE}
Almost surely, the Fibonacci drives of any order $m$ exhibit complete Hilbert-space ergodicity. 
\end{corollary}

{\it Proof sketch of Theorem \ref{th:01}}.%
\begin{figure*}
\centering
     \includegraphics[width=1\textwidth]{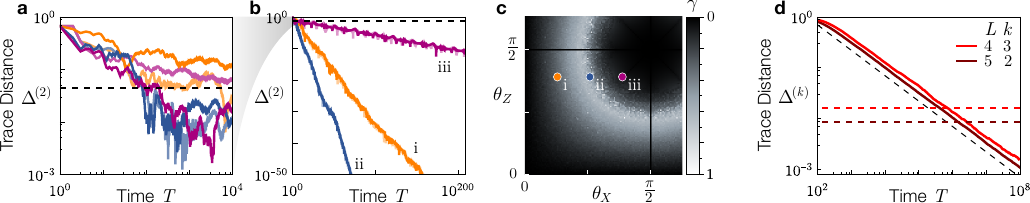}
     \caption{Numerical analysis of complete Hilbert-space ergodicity in the Fibonacci drive of order $1$. \textbf {a},\textbf{b}~Trace distance $\Delta^{(k=2)}(T)$ for single qubit rotations $A_0=e^{-i\theta_X{X}}$ and $A_1=e^{-i\theta_Z{Z}}$, for three pairs of angles $(\theta_X,\theta_Z)$: i~$(0.13\pi,0.39\pi)$ (orange), ii~$(0.26\pi,0.39\pi)$  (blue), and iii~$(0.39\pi,0.39\pi)$ (purple), and two initial states: $\ket{0}$ (darker), $\ket{+}$ (lighter). Lower bound $B(d=2)$ for time-independent evolution (black dashed). \textbf{c}~Power law exponent $\gamma$ as a function of $(\theta_X,\theta_Z)$ obtained by averaging $\Delta^{(2)}$ over  a pair of random initial states, for $T$ equal to the Fibonacci numbers $F_n$ up to $F_{3000}\approx 10^{626}$. \textbf{d}~$\Delta^{(k)}$  for many-body Fibonacci drive over a spin-$1/2$ chain of length $L$, averaged over 10 random product states $\ket{\psi}^{\otimes L}$ (solid lines). The bound $B(2^{L})$ is shown with horizontal dashed lines. A power law decay $\sim T^{-1/2}$ (black dashed) is provided as reference. 
     }
     \label{fig:fig3}
 \end{figure*}
The proof relies on the recursive and recurrent nature of the Fibonacci drives. Below we focus on the case $m=1$, and the more general case is proven in the SM~\cite{NoteSuppmat}.
To simplify notation, we drop the superscript $k$.

We note that it suffices to show that $\mathcal{N}_{\infty}\circ\mathcal{V}=\mathcal{N}_{\infty}$ with $\mathcal{V}[\,\cdot\,] \coloneqq\,V^{\otimes k}(\,\cdot\,)V^{\dagger\otimes{k}}$ for any $V\in\operatorname{SU}(d)$, where $\circ$ denotes the channel composition.
This is because $\mathcal{N}_\mathrm{Haar}$ is uniquely determined by a left or right invariance under any unitary rotation. 
This invariance is shown in three steps~\cite{NoteSuppmat}.

In the first step, we derive a recursive relation satisfied by $\mathcal{N}_T$ at times equal to a Fibonacci number $T=F_n$ \footnote{We use the convention $F_1=1$, $F_2=1$, and $F_n=F_{n-1}+F_{n-2}$}:
\begin{equation}
    {\mathcal{N}}_{F_{n}}=\frac{F_{n-1}}{F_{n}}{\mathcal{N}}_{F_{n-1}}+\frac{F_{n-2}}{F_{n}}{\mathcal{N}}_{F_{n-2}}\circ  {\mathcal{U}(F_{n-1})},
    \label{eqn:channel_recursive}
\end{equation}
where $\mathcal{U}(T)[\,\cdot\,]=U(T)^{\otimes k}(\, \cdot\,)U(T)^{\dagger\otimes k}$. 
Crucially, as long as $\mathcal{N}_\infty$ exists, one is free to choose any subsequence of times $(F_{n_\ell})_{\ell \in \mathbb{N}}$ to evaluate the limit at large $T$.
The key idea is to take an appropriate subsequence of times such that the relation in Eq.~\eqref{eqn:channel_recursive} reduces to the right invariance $\mathcal{N}_\infty=\mathcal{N}_\infty \circ \mathcal{L}$ under some unitary channel $\mathcal{L}[\,\cdot\,]=L^{\otimes k}(\, \cdot\,)L^{\dagger\otimes k}$,  where $L=\lim_{\ell \rightarrow \infty}U(F_{n_\ell})$.

In the second step, we show that there are at least two distinct subsequences such that $L=A_0$ and $A_1$, respectively.
This is enabled by the Poincar\'e Recurrence Theorem applied to a measure-preserving dynamical map 
$\Phi\colon(V,W)\mapsto(W,VW)$ for $V,W\in\operatorname{SU}(d)$, which generates our Fibonacci drive  $\Phi^{n}(A_1,A_0)=(U(F_n),U(F_{n+1}))$.
Specifically, the theorem states that almost any initial condition eventually comes arbitrarily close back to itself under sufficiently-long repeated application of $\Phi$. 
Translated to our case, almost surely $(A_1,A_0)$ is seen to be the limit of some subsequence of $(U(F_{n}),U(F_{n+1}))_{n \in \mathbb{N}}$. This immediately implies that $\mathcal{N}_\infty$ is right-invariant not only under the unitary channels $\mathcal{A}_0,\mathcal{A}_1$ generated by $A_0,A_1$ respectively, but also under any unitary channel $\mathcal{V}$ generated by arbitrarily many products of $A_0$ and $A_1$.

In the final step, we invoke a well-known fact in quantum information, that almost any pair $A_0,A_1$ of unitaries chosen uniformly from the Haar-measure generates the entire unitary group $\operatorname{SU}(d)$; 
 it is said that $A_0,A_1$ constitute universal ``gates''~\cite{Lloyd1995}. Taken together, this then implies  $\mathcal{N}_\infty$ is right-invariant under any unitary channel $\mathcal{V}$, as desired. \qed

Both Theorem \ref{th:01} and Corollary \ref{corollary:CHSE} demonstrate a strong form of ergodicity which quantum systems can exhibit: they illustrate how `disorder' (complete randomness) can emerge from `order' (a structured drive).
However, our analysis does not give us information about the speed at which $\rho_{T}^{(k)}$ converges to $\rho_{\mathrm{Haar}}^{(k)}$ for different choices of $A_0,A_1$ and initial state. 
To this end, we turn to numerical simulations.

{\it Numerical analysis.}---Below, we focus on the  Fibonacci drive of order $m=1$ and numerically compute the finite-time trace distance $\Delta^{(k)}(T)\coloneqq \tfrac{1}{2}\norm*{\rho_T^{(k)}-\rho_{\mathrm{Haar}}^{(k)}}_1$.

We first analyze the single qubit $d=2$ case, fixing $A_0=e^{-i\theta_X{X}}$ and $A_1=e^{-i\theta_Z{Z}}$ to be Pauli $X$ and $Z$ rotations by  angles $\theta_X$ and $\theta_Z$, respectively. Figures~\ref{fig:fig3}a and \ref{fig:fig3}b show $\Delta^{(2)}(T)$ for three choices of angles $(\theta_X,\theta_Z)$ indicated in Fig.~\ref{fig:fig3}c, starting from two initial states $\ket{0}$ and $\ket{+}=\left(\ket{0}+\ket{1}\right)/\sqrt{2}$. We observe an empirical power-law decay $\Delta^{(2)}\sim T^{-\gamma}$ going well-below the lower bound  $\sim 0.04$ for time-independent evolution in Eq.~\eqref{eq:boundtmindep}.  The power-law exponent $\gamma$ depends on the angles $(\theta_X,\theta_Z)$ as shown in Fig.~\ref{fig:fig3}c and on $k$ (see SM~\cite{NoteSuppmat}), but not on the initial state. Note that $\gamma=0$ at special points where $\theta_X=0,\pi/2$ or $\theta_Z=0,\pi/2$, signaling a breakdown of CHSE.
In such measure-zero cases, the set $\{A_0,A_1\}$ fails to generate all possible rotations. We present an in-depth study of CHSE for a single qubit in SM~\cite{NoteSuppmat}.

Finally, we study CHSE in a many-body system. 
We consider a spin-$1/2$ chain of length $L$ and generate the Fibonacci evolution with $A_0=e^{-iH_0 \tau}$ and $A_1=e^{-iH_1 \tau}$, where $H_0=\sum_{j=1}^{L}X_j+\sum_{j=2}^{L}X_{j-1}X_j+X_L/10$ and $H_1=\sum_{j=1}^{L}Z_j+\sum_{j=2}^{L}Z_{j-1}Z_j+Z_L/10$ for initial product states and $\tau=1$. 
In Fig.~\ref{fig:fig3}d we see again a power-law decay $\Delta^{(k)}\sim T^{-\gamma}$ with an exponent $\gamma\sim1/2$ for large $L$s and all $k$s simulated, which goes below the bound in Eq.~\eqref{eq:boundtmindep}. Understanding the origin of this seemingly universal exponent is an interesting future direction. Further note that both $H_0$ and $H_1$ are integrable Hamiltonians: it is the introduction of time dependence that brings non-trivial dynamical behavior \cite{Zvyagin2015, Zvyagin2018}: This suggests that CHSE can be achieved regardless of whether the system is integrable or non-integrable at any given fixed time.

{\it Discussion and outlook.---}%
There are several physical and conceptual implications of complete Hilbert-space ergodicity (CHSE).
CHSE essentially asserts that wavefunctions at sufficiently late times behave like random vectors in the Hilbert space.
This implies in turn that quantum many-body systems exhibiting CHSE can be rigorously shown to locally achieve thermalization to infinite temperature, since a typical Haar-random quantum state is highly entangled such that it locally appears maximally mixed~\cite{Popescu2006}.
By the same token, quantum systems with CHSE also {\it deep thermalize}: this is a recently developed notion of equilibration in which conditional states of a local subsystem, obtained via measurements of the complementary subsystem, achieve a maximally-entropic wavefunction distribution (on the local Hilbert space)~\cite{Claeys2022,Wilming2022,Ho2022,Ippoliti2022,Choi2023,Cotler2023,Ippoliti2023,Lucas2023,Shrotriya2023}. Using Theorem 2 of Ref.~\cite{Cotler2023}, we show that CHSE implies deep thermalization at late times~\cite{NoteSuppmat}.
We also present a numerical demonstration of deep thermalization in the Fibonacci drives.

Conceptually, CHSE has implications for the structure of solutions to time-dependent Schr\"odinger's equations. Under time-independent Hamiltonian dynamics, it is possible to decompose any initial state into a linear combination of energy eigenstates whose dynamics are trivial (specifically, having phases which wind in time), such that wavefunctions at arbitrarily late-times can be immediately recovered from the decomposition. The same is true for the dynamics of Floquet systems using  quasienergy eigenbases, which are guaranteed to exist by  Floquet's theorem. The ability to solve the time-dependent Schr\"odinger equation in terms of stationary solutions is called {\it reducibility} \cite{Murdock1978}. 
In time-quasiperiodic systems, the question of reducibility is nontrivial~\cite{Jauslin1991,Blekher1992}. Our results show that the time-quasiperiodic dynamics of the Fibonacci drives, or any other drives exhibiting CHSE, is irreducible, because the presence of quasienergy states is incompatible with the complete exploration of the Hilbert space in large-dimensional systems~\cite{Pilatowsky2024}. This opens the possibility that the computational complexity of solving quantum dynamics in such systems --- i.e., the computational resources required, necessarily grows unboundedly over time.

Our results lead to a number of interesting  future directions. First, systems exhibiting CHSE may yield settings in which rigorous studies on the growth of circuit complexity --- the minimal number of local gates needed to prepare a given quantum state from an unentangled state --- may be conducted.
Second, the question of how to experimentally verify and utilize CHSE, given finite coherence times of experiments, needs to be addressed.
Third, the relation between the breakdown of CHSE and integrability is an open question. We note that the presence of any conserved quantities --- local or even nonlocal, such as energy --- necessarily implies the breakdown of CHSE. The converse, however, calls for further investigation. 
To this end, it would be interesting to incorporate symmetries into the analysis. 
In particular, including the conservation of energy may bridge the gap between our dynamical notion of ergodicity and more conventional concepts based on the statistical properties of stationary states~\cite{Vikram2022}, ultimately providing a unified framework to understand quantum ergodicity.

\begin{acknowledgments}
We thank 
Y.~Bao,
A.~Chandran,
J.~Haferkamp,
J.~Jung,
D.~Long,
D.~Mark, and
I.~Marvian
for insightful conversations and acknowledge the MIT SuperCloud and Lincoln Laboratory Supercomputing Center, and Harvard University FAS Research Computing Center for providing computational resources utilized for our numerical analysis. This work is partly supported by NSF CUA (PHY-1734011), NSF CAREER (DMR-2237244), NSF STAQ (PHY-1818914) and the DARPA ONISQ program (W911NF201002). C.~B.~D. acknowledges support from the NSF through a grant for ITAMP at Harvard University. W.~W.~H. is supported by the NRF Fellowship, NRF-NRFF15-2023-0008, and the CQT Fellowship. 

\end{acknowledgments}

\bibliography{references}
\end{document}

\newcommand{\E}[0]{\mathop{{}\mathbb{E}}}
\newcommand{\SU}[1]{\operatorname{SU}#1}

\title{Supplemental Material: Complete Hilbert-Space Ergodicity in \\ Quantum Dynamics of Generalized Fibonacci Drives}
\author{Sa\'ul Pilatowsky-Cameo}
\affiliation{Center for Theoretical Physics, Massachusetts Institute of Technology, Cambridge, MA 02139, USA}
\author{Ceren Dag}
\affiliation{ITAMP, Harvard-Smithsonian Center for Astrophysics, Cambridge, MA 02138, USA}
\affiliation{Department of Physics, Harvard University, 17 Oxford Street Cambridge, MA 02138, USA}
\author{Wen Wei Ho}
\affiliation{Department of Physics, National University of Singapore, Singapore 117542}
\affiliation{Centre for Quantum Technologies, National University of Singapore, 3 Science Drive 2, Singapore 117543}
\author{Soonwon Choi}
\affiliation{Center for Theoretical Physics, Massachusetts Institute of Technology, Cambridge, MA 02139, USA}

\renewcommand{\thefigure}{S\arabic{figure}}
\renewcommand{\thetheorem}{S\arabic{theorem}}
\renewcommand{\thelemma}{S\arabic{lemma}}
\renewcommand{\theprop}{S\arabic{prop}}
\renewcommand{\theequation}{S\arabic{equation}}

\maketitle

In this supplemental material, we provide details on some of the claims made in the main text. Specifically, in Sec.~\ref{sec:lowerboundHam}, we derive a lower bound on the trace distance between the Haar average and the infinite-time average of any initial state undergoing quantum dynamics by a time-independent Hamiltonian, i.e., a ``no-go'' statement of complete Hilbert-space ergodicity (CHSE) in such systems.
In Sec.~\ref{app:cointossAbsoluteErgodicity}, we show that random quantum dynamics exhibits CHSE. 
In Sec.~\ref{app:quasiperiodic_form}, we demonstrate how the Fibonacci drives of any order $m$, introduced in the main text, can be understood as arising from a time-quasiperiodic drive with two mutually incommensurate frequencies.
In Sec.~\ref{app:inf_time_twirl}, we explain how the limit of the infinite-time averaging channels for a class of quasiperiodic systems related to the Fibonacci drives by an initial phase-shift almost always exists.
In Sec.~\ref{app:CHSE_Fib}, we present details of the proofs underlying Theorem 1 and Corollary  1, the main results of our paper, concerning the CHSE of the $m$-order Fibonacci drives.
In Sec.~\ref{app:numerics}, we explain the methods behind our numerical simulations.
In Sec.~\ref{app:qubit}, we present an in-depth numerical study of CHSE in the Fibonacci drive of a single qubit.
Lastly, in Sec.~\ref{app:deep_thermalization}, we showcase numerics demonstrating the emergence of deep thermalization in many-body Fibonacci drives.

\section{Non-ergodicity of time-independent Hamiltonian evolution}
\label{sec:lowerboundHam}
In this section we provide a bound to the degree of ergodicity achievable by time-independent Hamiltonian dynamics as quantified by Eq.~{(3)} of the main text. Specifically, we prove the following statement. Consider a time-independent Hamiltonian $H$, an arbitrary initial state $\ket{\psi}$, and define
\begin{align*}
    \rho^{(k)}_\infty&=\lim_{T\to \infty}\frac{1}{T}\int_0^T \dd{t} (e^{-i Ht}\dyad{\psi} e^{i Ht})^{\otimes k}.
\end{align*}
Using the trace norm $\norm{\rho}_1=\tr\sqrt{\rho \rho^\dagger}=\sum_{\lambda\in \text{Spec}(\rho)}\abs{\lambda},$ and defining $\Delta^{(k)}=\tfrac{1}{2}\norm*{\rho^{(k)}_{\mathrm{Haar}}-\rho_\infty^{(k)}}_1$, we have that 
for any $k\geq 2$, 
 \begin{equation}
 \label{eq:th1bound}
\Delta^{(k)}\geq B(d)\coloneqq \frac{1}{d+1}-\sqrt{\frac{1}{2d (d+1)}}.
 \end{equation}
This inequality implies that the time-independent Hamiltonian dynamics will never achieve CHSE, since $\rho_\infty^{\scriptscriptstyle(k)}$ is always distinct from $\rho_{\textrm{Haar}}^{\scriptscriptstyle(k)}$.

To show this, we first note that the trace distance is non-increasing under the application of a partial trace \cite[p. 407]{BookNielsen2010}, so for any $k\geq 2$ $$\Delta^{(k)}\geq\tfrac{1}{2}\norm{\tr_{k-2}\!\big(\rho^{(k)}_{\mathrm{Haar}}-\rho_\infty^{(k)}\big)}_1= \Delta^{(2)},$$ where $\tr_{k-2}(\,\cdot\,)$ denotes the partial trace over $k-2$ of the $k$ replicas of $\mathcal{H}$. Then, it is enough to lower bound $\Delta^{(2)}$. 
Second, we observe that the state $\rho_\mathrm{Haar}^{\scriptscriptstyle(k)}\coloneqq \mathop{\mathbb{E}}_{U\sim \mathrm{Haar}}\left[ ({U\dyad{\psi}U^{\dagger})^{\otimes k}}\right] 
$ can be analytically computed \cite{Harrow2013},
\begin{align}
\rho_\mathrm{Haar}^{(k)}=\frac{\sum_{\pi\in\mathcal{S}_k} \sum_{\alpha_1 \cdots \alpha_k}\dyad{\alpha_1 \alpha_2 \cdots \alpha_k}{\alpha_{\pi(1)}\cdots\alpha_{\pi(k)}}}{d(d+1)\cdots (d+k-1)},\label{eq:haarensemble}
\end{align} where $\ket{\alpha}$  is any orthonormal basis, labeled by $\alpha\in\{0,\dots,d-1\}$, and  $\mathcal{S}_k$ denotes the symmetric group of order $k$, that contains all possible permutations $\pi\colon\{1,2,\dots,k\} \to \{1,2,\dots,k\}$. If we take $\ket{\alpha}$ to be the eigenbasis of the Hamiltonian $H$, $H\ket{\alpha}=E_\alpha \ket{\alpha}$, and decompose $\ket{\psi}=\sum_{\alpha=0}^{d-1} c_\alpha \ket{\alpha}$, it is easy to see that
\begin{align}
\label{eq:rho:inf2}
 \rho_\infty^{(2)}
  =\sum_{{\alpha,\beta,\alpha',\beta'}} c_{\alpha}c_{\beta}c_{\alpha'}^*c_{\beta'}^*\dyad{\alpha\beta}{\alpha'\beta'} \delta_{ E_{\alpha}+E_{\beta},E_{\alpha'}+E_{\beta'}}.
\end{align}

If the Hamiltonian had no degeneracies in the spectrum, and also no degeneracies in the gaps of the spectrum (i.e., $E_\alpha-E_{\alpha'}=E_{\beta'}-E_{\beta} \implies (\alpha,\beta) =(\alpha', \beta')$ or $(\alpha,\beta) =(\beta', \alpha'))$  \footnote{This assumption is called the no 2-resonance condition \cite{Cotler2023}}, then the state given by Eq.~\eqref{eq:rho:inf2} would be completely dephased; the Kronecker-$\delta$  would only equal one if $(\alpha,\beta) =(\alpha', \beta')$ or $(\alpha,\beta) =(\beta', \alpha')$, leaving only a sum of terms with coefficients $|c_{\alpha}|^2|c_{\beta}|^2$, corresponding to the matrix entries  $\dyad{\alpha\beta}{\alpha\beta}$ and  $\dyad{\alpha\beta}{\beta\alpha}$. As we will see below, having only these matrix entries allows for analytical computation of $\Delta^{(2)}$, because these are also the nonzero matrix entries of $\rho_\mathrm{Haar}^{\scriptscriptstyle(2)}$. However, degeneracies  in the spectrum may exist, causing other non-dephased terms to remain. To account for this, we dephase possible degenerate subspaces via the application of a quantum channel, which cannot increase the trace distance. Specifically, we consider a (second order) dephasing channel 
$$\mathcal{D}[\rho^{(2)}]=\sum_{\alpha\beta}\dyad{\alpha\beta}\rho^{(2)}M(\alpha,\beta),$$
where 
$$M(\alpha,\beta)=\begin{cases}\dyad{\alpha\alpha}{\alpha\alpha},&\text{ if }\alpha=\beta \\
\dyad{\alpha\beta}{\alpha\beta} + \dyad{\alpha\beta}{\beta\alpha},&\text{ if } \alpha\neq \beta.\end{cases}$$
Applying this channel to Eq.~\eqref{eq:rho:inf2} yields $$\mathcal{D}[\rho_\infty^{(2)}]=\sum_{\substack{\alpha\beta}} \abs{c_{\alpha}}^2\abs{c_{\beta}}^2 M(\alpha,\beta).$$

Now, from Eq.~\eqref{eq:haarensemble} it follows that  $$\mathcal{D}[\rho_\mathrm{Haar}^{(2)}]=\rho_\mathrm{Haar}^{(2)}=\sum_{\alpha\beta}\frac{1+\delta_{\alpha\beta}}{d(d+1)}M(\alpha,\beta),$$
and then $$\mathcal{D}[\rho_\infty^{(2)}]-\mathcal{D}[\rho_\mathrm{Haar}^{(2)}]=\sum_{\alpha\beta} \Big(\abs{c_{\alpha}}^2\abs{c_{\beta}}^2-\frac{1+\delta_{\alpha\beta}}{d(d+1)}\Big)M(\alpha,\beta),$$ 
which can be easily diagonalized. Absolutely summing over all eigenvalues gives 
\begin{align*}
    \Delta^{(2)}&=\frac{1}{2}\norm{\rho_\mathrm{Haar}^{(2)}-\rho_\infty^{(2)}}_1\geq \frac{1}{2}\norm{\mathcal{D}[\rho_\mathrm{Haar}^{(2)}]-\mathcal{D}[\rho_\infty^{(2)}]}_1= \frac{1}{2}\sum_{\alpha\beta}\left|\abs{c_{\alpha}}^2\abs{c_{\beta}}^2-\frac{1+\delta_{\alpha\beta}}{d(d+1)}\right|,
\end{align*} 
where the inequality holds because the trace distance is non-increasing under the application of trace-preserving quantum channels \cite[p. 406]{BookNielsen2010}. Considering only the terms with $\alpha=\beta$ yields
$$\Delta^{(2)}\geq \frac{1}{2}\sum_{\alpha} \left|\abs{c_{\alpha}}^4-\frac{2}{d(d+1)}\right|.$$ 
This may be shown to be lower bounded by $B(d)$, as defined in Eq.~\eqref{eq:th1bound}, by inserting $\xi=\sqrt{{2}/{d(d+1)}}$ into the following lemma.

\begin{lemma}
\label{lemmF}
Let $C=\{(a_1, \dots, a_d)\in[0,1]^d \,|\, \sum_n a_n=1\},$ and consider $F:C\to \mathbb{R}$ given by
$$F(a_1, \dots, a_d)= \sum_{n=1}^d \left|a_{n}^2-\xi^2\right|,$$
for any $\xi\in[1/d,1]$. Then, for all $\bm a\coloneqq (a_1,\dots,a_d)\in C$,
$$F(\bm a)\geq \xi^2\left(d-\frac{1}{\xi}\right).$$
\end{lemma}
\begin{proof}
We only require elementary calculus techniques. Consider the minimum $\bm{a}_0=(a_1,\dots,a_d)\in C$ of $F$. This point exists because $F$ is continuous and $C$ is compact.   It suffices to show that $F(\bm a_0)\geq \xi^2\left(d-\xi^{-1}\right)$.  In the following two claims, we will show that for all $n$ except at most one, $a_{n}=0$ or $a_{n}=\xi$. Let $\delta_n=a_{n}^2-\xi^2$, so that $\delta_n=0$ means $a_{n}=\xi$.

\noindent\underline{Claim 1}: If $\abs{\delta_m}, \abs{\delta_n}>0$ and $a_m>0$ then either $a_n=0$ or $\delta_n=\delta_m> 0$.

\begin{list}{\it{ Proof.}}{\setlength\topsep{0pt}}
\item Consider \begin{equation}
\label{eq:equilibriumfunction}
    g(\epsilon) = \left|(a_{m}+\epsilon)^2-\xi^2\right|+\left|(a_{n}-\epsilon)^2-\xi^2\right|.
\end{equation}
Both absolute values are positive at $\epsilon=0$, so for small enough $\abs{\epsilon}$, 
\begin{align*}
    g(\epsilon)&=\text{sign}(\delta_m)\left((a_{m}+\epsilon)^2-\xi^2\right)+\text{sign}(\delta_n)\left((a_{n}-\epsilon)^2-\xi^2\right)
\end{align*}
and $g$ is differentiable at $\epsilon=0$. Assume that $a_n>0$. Then for any small enough $\abs{\epsilon}$, $\bm a_\epsilon \coloneqq \bm a_0  +{\epsilon} \bm (\bm e_m -\bm e_n)\in C$ ($\bm{e}_j$ are the canonical vectors in $\mathbb{R}^d$), and $F(\bm a_\epsilon)=g(\epsilon)$, so $g'(0)$ is the directional derivative of $F$ in the $\bm e_m -\bm e_n$ direction, at $\bm a_0$. Because $\bm a_0$ is the global minimum, then $g'(\epsilon)=0$. 
Computing the derivative,
\begin{align*}
    0=g'(\epsilon=0) =2(\text{sign}({\delta_m)}a_{m}-\text{sign}({\delta_n)}a_n),
\end{align*} so $\text{sign}({\delta_m)}a_{m}=\text{sign}({\delta_n)}a_n$. Because, they must be nonnegative,  $a_n=a_m$ and $\delta_n=\delta_m$. By the second derivative criterion $0\leq g''(0)=2(\text{sign}({\delta_n)} + \text{sign}({\delta_m)})$, so $\delta_n=\delta_m>0$.
$\blacksquare$
\end{list}

\noindent\underline{Claim 2}: For all $n$, $\delta_n\leq 0$.
\begin{list}{\it{ Proof.}}{\setlength\topsep{0pt}}\item First note that there exists at least one $\delta_m\leq 0$.
If not, we would have $a_m^2>\xi^2$ for all $m$, and summing over all $m$ we would obtain a contradiction $1=\sum_m a_m >d\xi\geq 1$.
Now, assume by contradiction that there exists $\delta_n>0$. Define $g$ as in Eq.~\eqref{eq:equilibriumfunction}. In this case $g$ might not be differentiable at $\epsilon=0$, because $\delta_m$ might be zero, but still $g$ is semi-differentiable.  The right derivative $g_+'(0)$ is the (right) directional derivative of $F$ in the $\bm e_m -\bm e_n$ direction, at $\bm a_0$. Because $\bm a_0$ is the global minimum, then $g_+'(0)\geq0$. For small enough $\epsilon>0$, $$g(\epsilon)=-\left((a_{m}+\epsilon)^2-\xi^2\right)+\left((a_{n}-\epsilon)^2-\xi^2\right),$$ so $0\leq g_+'(\epsilon=0)=2(-a_{m}-a_n)$, which implies $a_m=a_n=0$. This contradicts $\delta_n>0$. $\blacksquare$ \end{list}

From Claims 1 and 2 it follows that there is at most one $n_*$ with $\xi>a_{n_*}>0$, and the remaining $a_n$ are either $a_n=0$ or $a_n=\xi$. Let $M$ be the number of $a_n$ which are equal to $\xi$. By normalization, if there is $\xi> a_{n_*}>0$, then $a_{n_*}=1-\xi M$. If not, it must be that $\xi M=1$. In both cases $M\leq 1/\xi$ and
\begin{align}
\label{eq:fminintermsofM}
    F(\bm a_0) &=\xi^2(d-M)-\left(1-\xi M\right)^2 =\xi^2\left(d-\frac{1}{\xi}+\frac{1}{4}-\left(\frac{1}{\xi}-M-\frac{1}{2}\right)^2\right). \nonumber
 \end{align}
\newcommand{\floor}[1]{\left\lfloor #1 \right\rfloor}
\newcommand{\ceil}[1]{\left\lceil #1 \right\rceil}

To finally show that $F(\bm a_0)\geq \xi^2\left(d-1/{\xi}\right)$, we just have to verify that $1/4 - (1/\xi-M-1/2)^2 \geq 0$. Indeed, let us check that $0\leq \frac{1}{\xi} - M<1$ (i.e., $M=\floor{\frac{1}{\xi}}$). We have already seen that $M\leq 1/\xi$. 
For the other inequality, by normalization, note that if there exists $\xi>a_{n_*}>0$, then $1=M\xi + a_{n_*}< (M+1)\xi$, and if not,  $1=\sum_na_n=M\xi<  (M+1)\xi$. In any case $M+1>1/\xi$, which completes the proof.
\end{proof}

\section{Complete Hilbert-space ergodicity by flipping a coin}
\label{app:cointossAbsoluteErgodicity}
In this section, we show that CHSE is achieved almost surely by randomly applying a pair of unitaries according to the outcomes of sequence of coin flips. Specifically, consider two unitaries $A_0$ and $A_1$. At at each time $n\in\mathbb{N}$, flip a coin, labeled by $c_n\in \{0,1\}$, and apply the corresponding unitary $A_{c_n}$. This generates the evolution 
$$\ket{\psi(n)}=A_{c_n}A_{c_{n-1}}\cdots A_{c_1}\ket{\psi(0)},$$ which we show to satisfy CHSE, as long as $A_0$ and $A_1$ generate the entire $\SU(d)$.

\begin{theorem}
    Let $A_0$ and $A_1$ form a computationally complete  set of `gates', meaning that the group generated by $\{A_0,A_1\}$ is dense in $\SU(d)$. Consider an infinite binary sequence $(c_n)_{n\in \mathbb{N}}$, where $c_n\in \{0,1\}$ are independent identically-distributed random variables, sampled from a distribution with support on both $0$ and $1$. Then the quantum system with evolution operator $U(n)=A_{c_n}A_{c_{n-1}}\cdots A_{c_1}$, and $U(0)=\mathds{1}$ satisfies CHSE with probability $1$, over the realization of the sequence $c_1c_2c_3\cdots$.
\end{theorem}
\noindent Note that we are not considering an ensemble over different choices of $A_0$ and $A_1$; these unitaries remain fixed. The set of events is taken over the realization of the random binary sequence $c_1c_2c_3\cdots$ in time, i.e., over the infinite sequence of coin tosses. A single realization of this random sequence is drawn, and used to generate the quantum evolution.
Thus, an event in which CHSE is not achieved is if $c_{2n} = 1$ and $c_{2n+1} = 0$, for example, so that $U(2n) = (A_1 A_0)^n$, i.e., a Floquet drive. Such events are, however, of zero measure. 
Furthermore, note that the coin is allowed to be partially biased, i.e., the distribution describing the binary numbers $c_n$ not need to be uniform, just needs to have support on both outcomes.
\begin{proof}
In summary, the sequence $U(n)$ generates a random walk on the compact group $\text{SU}(n)$, and these random walks are known to be ergodic \cite{Emerson2005,Berger2012}, with unit probability. Explicitly, we use the following result.

{\noindent\bf Corollary 3.1 of \cite{Berger2012}. }{\it Let $(\xi_n)_{n\in \mathbb{N}}$
be an independent identically-distributed sequence in the metrizable compact group $G$ with the common distribution having support $S$. Then, the sequence $(\xi_1\cdots\xi_n)_{n\in \mathbb{N}}$ is, with probability one, uniformly distributed in $G$ if and only if $\bigcup_{n\in\mathbb{N}} S^n$ is dense in $G$.}

\noindent That the sequence is {\it uniformly distributed} in the group $G$ means that for any continuous real function $f\colon G\to \mathbb{R}$, 
$\lim_{N\to \infty}\frac{1}{N}\sum_{n=0}^{N-1} f(\xi_1\cdots\xi_n)=\E_{\xi \sim \mathrm{Haar}}[f(\xi )]$. The notation $S^n$ represents the set of all group elements of the form $s_1s_2\cdots s_n$, with $s_j\in S$.

In our case, we take $\xi_n=A_{c_n}$,  $G=\SU(d)$, $\xi_1\cdots\xi_n=U(n)$ and $S=\{A_0,A_1\}$, which satisfies that $\bigcup_{n\in\mathbb{N}} S^n$ is dense in $G$ because $\{A_0,A_1\}$ forms a computationally complete set of gates. Then we obtain that for any continuous function $f\colon\SU(d)\to \mathbb{R}$,
$$\lim_{N\to \infty}\frac{1}{N}\sum_{n=0}^{N-1} f(U(n))=\E_{U\sim \mathrm{Haar}}[f(U)].$$ Taking the functions $f(U)$ to be the real and imaginary part of each entry of the matrix $(U\dyad{\psi(0)}U^\dagger)^{\otimes k}$,
we obtain CHSE
\begin{equation*}
    { \lim_{N\to \infty}\frac{1}{N}\sum_{n=0}^{N-1} \dyad{\psi(n)}^{\otimes k }=\rho_{\mathrm{Haar}}^{(k)}}. \label{app:absergapp} \qedhere
\end{equation*}
\end{proof}

\section{Quasiperiodic form of Fibonacci drives}
\label{app:quasiperiodic_form}
We show here that the Fibonacci drives of order $m$ can be generated by a time-quasiperiodic Hamiltonian, i.e., where the time dependence can be written in terms of two angles linearly wrapping around the torus. Specifically, in continuous time $t\in [0,\infty)$, the Fibonacci drive of order $m$ is defined by the unitary evolution operator $U(0\leq t<1)=\mathds 1$ and
\begin{align}
\label{eq:Uoperatorconttime}
  U(t\geq 1) = A_{\omega_{\lfloor t\rfloor}}  \cdots  A_{\omega_3}A_{\omega_2}A_{\omega_1},
\end{align}
for two fixed unitaries $A_0,A_1\in \SU(d)$, where $\omega_n$ is the $n$th symbol in the Fibonacci word of order $m$, $W_\infty$, which is generated by infinitely applying the repeated concatenation $W_{j}=W_{j-1}^m W_{j-2}$ starting from $W_0=1$ and $W_1=0$. This unitary evolution is generated by a sequence of kicks at integer times, given by the Hamiltonian
\begin{equation}
\label{eq:eq_kicking_Hamiltonian}
    H(t)=\sum_{n=1}^\infty\delta(n-t)H_{\omega_n},
\end{equation}
where $e^{-iH_0}=A_0$ and $e^{-iH_1}=A_1$.
We will show that $H(t)$ is quasiperiodic, meaning that its time dependence can be written in terms of two angles \begin{equation}\label{eq:anglestheta1theta2}\theta_1(t)=2\pi t \mod 2\pi,\quad\quad\theta_2(t)=\Omega t \mod 2\pi,\end{equation} 
where  \begin{equation}
    \Omega= \frac{\pi}{ m}\left(2+m-\sqrt{m^2+4}\right)\in(0,2\pi).
    \label{eq:angular_speed_eq}
\end{equation}Explicitly, one can write
\begin{equation}
\label{eq:QuasiperiodicHamiltonian}
H(t)=\delta(\theta_2(t))\times\begin{cases} 0 &\text{if }\theta_1(t)=0\\H_0&\text{if } \theta_1(t)\in [0, 2\pi-\Omega]
    \\H_1&\text{if } \theta_1(t)\in (2\pi-\Omega,2\pi).\end{cases}
\end{equation}

Figure 2c of the main text shows the Hamiltonian \eqref{eq:QuasiperiodicHamiltonian} in the torus, for $m=1$. The circle $\theta_2=0$ is colored orange if $\theta_1\in [0,2\pi-\Omega]$ and blue if $\theta_1\in (2\pi-\Omega,2\pi)$. The line $(\theta_1(t),\theta_2(t))$ given by Eq.~\eqref{eq:anglestheta1theta2} is drawn in black. When $\theta_2(t)=0$, this black line crosses the colored circle, and the state gets kicked with $H_0$ or $H_1$, depending on the value of $\theta_1(t)$.

To see why Eq.~\eqref{eq:QuasiperiodicHamiltonian} is true, we will give a brief introduction to the theory of the so-called Sturmian words (SWs). The standard Sturmian words (SSWs) are those infinite sequences of zeros and ones $W_\infty=\omega_1\omega_2\omega_3\cdots$  generated by the repeated concatenation rule 
\begin{align}
\label{eq:repeated_concat}
    W_{j+1}=W_{j}^{q_{j}} W_{j-1} \quad\quad(j\geq 1),
\end{align}
starting from $W_0=1$ and $W_1=0$, where  $q_j$ are non-negative integers with $q_1>0$. The Fibonacci words of order $m$ are examples of SSWs obtained by setting $q_1=q_2=\cdots=m$.

Another equivalent way to construct a SSW is by  \textit{coding an irrational rotation}. The idea is to partition the circle $\mathbb{T}=[0,2\pi)$ in two disjoint subintervals $I_0 =[0,2\pi-\Omega]$  and $I_1 =(2\pi-\Omega,2\pi)$, for some angle $\Omega\in (0,2\pi)$, and consider the rotation $R\colon \mathbb{T}\to \mathbb{T}$ given by $R(\theta)=\theta+ \Omega \mod 2\pi$. To construct the  Sturmian word $W_\infty=\omega_1\omega_2\omega_3\cdots$, start at some $\theta_0\in \mathbb{T}$, and see whether  $R(\theta_0)\in I_0$ or $R(\theta_0)\in I_1$, writing, $\omega_1=0$ or $\omega_1=1$, respectively. Then do the same for $R^2(\theta_0)=R(R(\theta_0))$. Repeating this process $n$ times, we choose $\omega_n=0$ or $\omega_n=1$ depending on whether $R^n(\theta_0)$ falls in $I_{0}$ or $I_{1}$. In symbols,
\begin{equation}
\label{eq:irrational_rotation}\omega_n=\begin{cases}
    0 &\text{if }\text{mod}_{2\pi}( n\Omega + \theta_0) \in [0,2\pi-\Omega] \\1&\text{if } \text{mod}_{2\pi}( n\Omega+ \theta_0)\in (2\pi-\Omega,2\pi), \end{cases}
\end{equation}
where $\text{mod}_{2\pi}( n \Omega +\theta_0)$ gives the remainder of the number $ n\Omega + \theta_0$ after dividing by $2\pi$. 

The Sturmian word $\omega_1\omega_2\omega_3\dots$ is called non-standard if $\theta_0\neq 0$. If $\theta_0=0$, it turns out that Eq.~\eqref{eq:irrational_rotation} generates a SSW, as defined by Eq.~\eqref{eq:repeated_concat}, where the numbers $q_1,q_2,q_3,\dots$ are related to $\Omega$ by the following continued fraction (see \cite[p. 50]{DeLuca1997} and \cite[6.1.16]{BookPytheasFogg2006})
\begin{equation}
    \frac{\Omega}{2\pi}=1-\cfrac{1}{1+\cfrac{1}{q_1+\cfrac{1}{q_2+\cfrac{1}{q_3+\ddots  } } }}.\label{eq:cont_fraction}
\end{equation}
The Fibonacci word of order $m$ is obtained by taking all $q_j$ equal to $m$ in Eq.~\eqref{eq:repeated_concat}. Inserting this into the continued fraction \eqref{eq:cont_fraction} gives $\Omega=2\pi[1-1/(1+1/\mu)]$, where $\mu=\tfrac{1}{2}\left(m+\sqrt{m^2+4}\right)$ is the $m$th metallic ratio, which for $m=1$ is the golden ratio, for $m=2$ is the silver ratio, and so on. This value of $\Omega$ gives Eq.~\eqref{eq:angular_speed_eq}, and Eq.~\eqref{eq:QuasiperiodicHamiltonian} is obtained by substituting Eq.~\eqref{eq:irrational_rotation} into Eq.~\eqref{eq:eq_kicking_Hamiltonian}.

\section{``Almost-existence'' of infinite-time averaging channel of Fibonacci drives}
\label{app:inf_time_twirl}
As we stated in the main text, we could not guarantee the existence of the infinite-time averaging channel (ITAC) $\mathcal{N}^{\scriptscriptstyle (k)}_\infty$ [see below Eq.~(5) of the main text] for the Fibonacci drives, which are defined using a standard SW, and had to assume their existence for our proofs. 
However, interestingly, we {\it can} prove the limit exists for drives defined by non-standard SWs,  where an initial phase $\theta_0$ in the coding of the irrational rotation is allowed, for almost all $\theta_0$.

To see this, let us first recast the dynamics under the Fibonacci drives into a useful, alternative form. 
The quasiperiodicity of the Fibonacci drives allows to write the evolution operator $U(t)$ at integer times $t=n$ as the image of an iterated map. As before, consider the rotation of the circle $R(\theta)=\theta+\Omega \mod 2\pi$ and denote
$$\chi(\theta)=\begin{cases} 0&\text{if } R(\theta)\in [0, 2\pi-\Omega]
    \\1&\text{if } R(\theta)\in (2\pi-\Omega,2\pi)\end{cases}.$$
The map given by \begin{equation} \label{eq:autonomous_map_Fibonacci}
    P(\theta,V)=(R(\theta),A_{\chi(\theta)}V)
\end{equation} is called a skew-product dynamical system. It is easy to see that $P^n(0,\mathds{1})=(R^n(0),U(n))$. That is, the Fibonacci evolution operator $U(n)$ can be recovered by applying $P$ a total of $n$ times to the initial point $(\theta_0=0, V_0=\mathds{1})$. 

Consider next quantum dynamics generated by $U(t)$ as in Eq.~\eqref{eq:Uoperatorconttime}, but now with $\omega_n$ defined by the coding of an irrational rotation \eqref{eq:irrational_rotation},  allowing for an initial phase $\theta_0$ (i.e., allowing for non-standard SWs). Then $P^n(\theta_0,V_0)=(R^n(\theta_0),U(n)V_0)$ for any $V_0\in \SU(d)$ and $n\in\mathbb{N}$.
Birkhoff's Ergodic Theorem (BET) guarantees that for almost any initial angle $\theta_0$, the infinite-time averages over the quantum dynamics generated by $U(t)$ exist \cite{Cornfeld1982}. This is because the ITAC for $U(t)$ can be written as the Birkhoff average $\mathcal{N}_\infty^{\scriptscriptstyle (k)}=\lim_{T\to \infty} \frac{1}{T}\sum_{t=0}^{T-1} f(P^n(\theta_0,V_0=\mathds{1}))$, where $f$ maps $(\theta,V)$ to the unitary quantum channel $(\,\cdot\,)\mapsto V^{\otimes k}(\, \cdot\,)V^{\dagger\otimes k}$, regardless of the value of $\theta$. The BET guarantees that these averages exist for almost any choice of initial $(\theta_0,V_0)$. We may get rid of the initial $V_0$ by performing an initial rotation by $V_0^\dagger$, guaranteeing the existence of the ITAC for almost any $\theta_0$. 

Unfortunately, the argument above does not imply that the ITAC exists for the Fibonacci drives of our work, which have the particular initial angle $\theta_0=0$. For that, one would have to show that $0$ generally does not fall in the zero-measure set of angles $\theta_0$ for which the Birkhoff averages may fail to converge. This is an interesting problem which we leave open.

\section{Detailed proof of complete Hilbert-space ergodicity in the Fibonacci Drive}
\label{app:CHSE_Fib}
In this section, we present the full proof of our main result, Theorem 1 of the main text, which, for convenience, we restate here.

    {\noindent\bf Theorem 1 }(Restatement). {\it
    If $ A_0, A_1$ are chosen at random from $\operatorname{SU}(d)$, with unit probability, the ITAC (assumed to exist) of the Fibonacci drive of order $m$ satisfies
     \begin{align}   \forall k\in\mathbb{N}:\quad  \mathcal{N}^{(k)}_\infty=\mathcal{N}_{\mathrm{Haar}}^{(k)},
     \end{align}
     where $\mathcal{N}_{\mathrm{Haar}}^{(k)}[\,\cdot\,] = \int dU U^{\otimes k} ( \,\cdot\, ) U^{\dagger \otimes k}$ is the $k$-fold Haar-averaging channel over $\operatorname{SU}(d)$.
 }

 \noindent We will follow the same structure as the sketch of proof for $m=1$ presented in the main text. We will split the argument into four lemmas. As in the main text, we will henceforth drop the superscript $(k)$ in our notations. 

Any unitary $U\in \SU(d)$ generates a unitary channel given by $\mathcal{U}[\,\cdot\,] \coloneqq U^{\otimes k}(\, \cdot\,)U^{\dagger\otimes k}$. Because the Haar-averaging channel is the only channel which is right-invariant under any unitary rotation, we only need to show that   \begin{align}
\label{eq:setAequivance}
        \Pr[ \forall{k\in \mathbb{N}},\, \forall{U\in \SU(d)}:   \mathcal{N}_\infty\circ \mathcal{U}=\mathcal{N}_\infty]=1.
    \end{align}
    To  find unitaries $U$ for which the invariance $\mathcal{N}_\infty\circ \mathcal{U}=\mathcal{N}_\infty$ holds, we will show some recursive relations that occur at special times.

Let $S_n$ denote the sequence of numbers generated by the recurrence relation \begin{align} 
\label{eq:SnRecurrence}
S_0=S_1=1,&& S_n=m S_{n-1}+S_{n-2}
\end{align}
for $n\geq 2$. The case $m=1$ gives the Fibonacci numbers, $F_n=S_{n-1}$; we may refer to the general case $m \geq 2$ as ``generalized Fibonacci numbers''.  When evaluated at $t=S_n$, the evolution operator $U(S_n)$ follows a recursive relation %
    \begin{align}\label{eq:recursiveForumlaoperator}
        U(S_1)&=A_0, &
        U(S_2)&=A_1 A_0^m, &
        U(S_{n+1}) &= U(S_{n-1})U(S_{n})^m,& 
    \end{align}
    for $n\geq 2$. This follows directly from the recursive construction of the Fibonacci words. Moreover, the TAC evaluated at times $T=S_n$ 
     \begin{equation}
    \label{eq:tchannel}
     \mathcal{N}_{{S_{n}}}=   \frac{1}{S_{n}}\sum_{t=0}^{S_{n}-1} \mathcal{U}(t),
    \end{equation}
    is generated by the following recursive rule:
    \begin{align}
\label{eq:recursiveForumla}{\mathcal{N}}_{S_0}={\mathcal{N}}_{S_1}&=\text{Id} \quad \text{(identity channel)}, &
        {\mathcal{N}}_{S_{n+1}} &=\sum_{y=0}^{m}\frac{S_{n-\delta_{ym}}}{S_{n+1}}\,\mathcal{N}_{S_{n-\delta_{ym}}}\circ\mathcal{U}(S_{n})^{\circ y},&
    \end{align}
   for $n\geq 1$, where $(\,\cdot\,)^{\circ y}=(\,\cdot\,)\circ (\,\cdot\,)\circ \cdots\circ  (\,\cdot\,)$ ($y$ times), $(\,\cdot\,)^{\circ 0}=\text{Id}$, and $\delta_{ym}$ is the Kronecker delta.
   
   \begin{proof}[Proof of Eq.~\eqref{eq:recursiveForumla}]
    One can check the case $n=1$ directly:
    $$\mathcal{N}_{S_2}=\sum_{y=0}^{m}\frac{1}{S_{2}}\,\mathcal{U}({S_1})^{\circ y}=\frac{1}{S_{2}}\sum_{t=0}^{S_2-1}\,{\mathcal{U}(t)}.$$
    For the rest of the proof assume $n\geq 2$. For each $y\in\{0,1\dots,m\}$, denote $T_{nmy}=S_{n-\delta_{ym}}$.
    
    \noindent\underline{Claim}: For any integer time $0\leq t< T_{nmy}$, the evolution operator $U(t)$ satisfies   \begin{equation*}U(t) U(S_{n})^y= U(t+y S_{n}).\end{equation*}
\begin{list}{\it{ Proof.}}{\setlength\topsep{0pt}}
    \item 
   For integer times $t_1 > t_0\geq 0$ denote $$U(t_1;t_0)=A_{\omega_{t_1}}A_{\omega_{t_1-1}}\cdots A_{\omega_{t_0+1}},$$ 
   and $U(t_0;t_0)=\mathds{1}$.
   This is the evolution operator from time $t_0$ to time $t_1$. With this notation $U(t)=U(t;0)$, so \begin{align*}
    \quad\quad U(t+yS_n)=&\, U(t+yS_n;yS_n)U(yS_{n};(y-1)S_n)U((y-1)S_{n};(y-2)S_n)\cdots U(S_{n};0).
   \end{align*} One can check that $U(xS_n;(x-1)S_n)=U(S_n)$ for all $x\in\{1,2,\dots,y\}$ and $U(t+yS_n;yS_n)=U(t)$ if $0\leq t< T_{nmy}$ because of the recursive structure of the Fibonacci word of order $m$, whose entries satisfy 
   $\omega_{j}=\omega_{j+yS_n}$ for all $1\leq j\leq S_{n-\delta_{ym}}=T_{nmy}$. $\blacksquare$
 \end{list}
 
 From the claim, it follows that $\mathcal{U}(t)\circ \mathcal{U}(S_n)^{\circ y}=\mathcal{U}(t+yS_n)$ whenever $0\leq t< T_{nmy}$, so, inserting ${S_{n-\delta_{ym}}}\,\mathcal{N}_{S_{n-\delta_{ym}}}=\sum_{t=0}^{{\smash{T_{nmy}-1}}} \mathcal{U}(t)$, we obtain
        \begin{align*}
  &\sum_{y=0}^{m}{S_{n-\delta_{ym}}}\,\mathcal{N}_{n-\delta_{ym}}\circ\mathcal{U}(S_{n})^{\circ y} =\sum_{y=0}^{m} \quad\sum_{t=0}^{\mathclap{{T_{nmy}-1}}} \mathcal{U}(t+yS_n)=\sum_{t=0}^{S_{n+1}-1} \mathcal{U}(t)=S_{n+1}\mathcal{N}_{n+1}.
    \end{align*}
    where we used $S_{n-1}+mS_n=S_{n+1}$ in the second equality.
    This is precisely Eq.~\eqref{eq:recursiveForumla}.      \end{proof}

Recall that we are interested in showing that, with unit probability, $\mathcal{N}_\infty\circ \mathcal{U}=\mathcal{N}_\infty$ for all $U\in \SU(d)$. Using the recursive formula \eqref{eq:recursiveForumla}, we can show that this is true for a particular class of unitaries. The sequence generated by evaluating the evolution operator at times $S_n$, $(U(S_{n}))_{n\in \mathbb{N}}$ does not necessarily converge, but by compactness it has subsequences that do. The ITAC remains invariant  under right-composition by the limits of these subsequences.

\begin{lemma}[Invariance under subsequential limits]
\label{lem:subsequentiallimitsinvariant}
    For any limit of a subsequence of $U(S_n)$,  $L=\lim_{l\to \infty}U(S_{n_l})$, the ITAC satisfies   $\mathcal{N}_\infty=\mathcal{N}_\infty\circ \mathcal{L},$ where  $\mathcal{L}$ represents the unitary channel $(\,\cdot\,)\mapsto L^{\otimes k}(\,\cdot\,) L^{\dagger\otimes k}$.
    
\end{lemma}
\begin{proof}

Substituting the subsequence $n_l$ and taking the limit $l\to \infty$ on Eq.~\eqref{eq:recursiveForumla}, we obtain
\begin{equation}
\label{eq:chanlimmetratioblah}
\mathcal{N}_{\infty}=\sum_{y=0}^{m}\mu^{-1-\delta_{ym}}\, \mathcal{N}_{\infty}\circ \mathcal{L}^{\circ y},
\end{equation}
where $ \mu=\lim_{n\to \infty}{S_{n+1}}/{S_n}= \tfrac{1}{2}\left(m+\sqrt{m^2+4}\right)$  is the $m$'th metallic ratio, which satisfies $\mu^2-m\mu=1$. Note that, in taking this limit, we are making use of our overall assumption that $\mathcal{N}_\infty$ indeed exists.
Equation~\eqref{eq:chanlimmetratioblah} can then be turned into
\begin{equation}
\label{eq:anyxinv}
\mathcal{N}_\infty=\mathcal{N}_\infty\circ\sum_{y=1}^{m}c_y \mathcal{L}^{\circ y},
\end{equation}
where $c_y=(1+\delta_{ym}(\mu-m-1))/(\mu-1)$. Note that this completes the proof for $m=1$, as in this case $c_1=1$.

Assume $m\geq 2 $. We will construct a sequence of vectors $\boldsymbol a_n=(a_n^{(1)},\dots, a_n^{(m)})^\intercal$ such that
\begin{equation}
\mathcal{N}_\infty- \mathcal{N}_\infty\circ \mathcal{L}= \sum_{y=1}^{m}a_n^{(y)}\mathcal{N}_\infty\circ \mathcal{L}^{\circ n+y-1 }\label{eq:diffseqeq}
\end{equation}
and show that  $\lim_{n\to \infty} {\boldsymbol a}_n ={\boldsymbol 0}$, which will complete the proof. Evidently, for $n=0$,  Eq.~\eqref{eq:diffseqeq} holds with $\boldsymbol a_0=\left(1, -1,0,\dots,0\right)^\intercal$. Now, from Eq.~\eqref{eq:anyxinv}, by right-composing with $\mathcal{L}^{\circ n}$ we obtain
\begin{equation*}
\mathcal{N}_\infty\circ \mathcal{L}^{\circ n}= \mathcal{N}_\infty\circ\sum_{y=1}^{m}c_y \mathcal{L}^{\circ n+y}.
\end{equation*}
Thus,
\begin{align*}
\mathcal{N}_\infty\circ\sum_{y=1}^{m}a_n^{(y)}\mathcal{L}^{\circ n+y-1 }&=\mathcal{N}_\infty\circ\Big(a_{n}^{(1)}\sum_{y=1}^{m}c_y \mathcal{L}^{\circ n+y}+\sum_{y=2}^{m}a_n^{(y)}\mathcal{L}^{\circ n+y-1 }\Big)\\&=\mathcal{N}_\infty\circ\sum_{y=1}^{m}\Big(a_n^{(1)}c_y+(1-\delta_{ym})a_n^{(y+1)}\Big)\mathcal{L}^{\circ n+y}=\mathcal{N}_\infty\circ\sum_{y=1}^{m}a_{n+1}^{(y)}\mathcal{L}^{\circ n+y },
\end{align*}
where we define
\begin{align}
    {\boldsymbol a}_{n+1}=\left(\frac{a_n^{(1)}}{\mu-1}+a_n^{(2)},\dots,\frac{a_n^{(1)}}{\mu-1}+a_n^{(m)},a_n^{(1)}\frac{\mu-m}{\mu-1}\right)^\intercal\nonumber
\end{align}
for $n\geq 0$.
This sequence can be obtained from iterated matrix multiplication ${\boldsymbol a}_{n}=M^n{\boldsymbol a}_{0}$, where

$$M=\begin{pmatrix} 
             \tfrac{1}{\mu-1} &    1    &    0    &      0    & \cdots   & 0\\
             \tfrac{1}{\mu-1} &    0    &    1    &     0    &  \cdots    &  0 \\
             \tfrac{1}{\mu-1} &    0    &    0     &    1    &  \cdots    &  0 \\
                 \vdots      & \vdots  &     \vdots  &     & \ddots &  \vdots \\      
             \tfrac{1}{\mu-1} &    0    &    0   &    0    & \cdots   & 1\\
\tfrac{\mu-m}{\mu-1}&    0    &    0     &    0 &  \cdots      & 0
     \end{pmatrix}.$$

 One can verify that the matrix $M$ is primitive, meaning that all its entries are non-negative and all the entries of one of its powers ($M^m$) are positive. Further, the positive vector $\bm{w}=(1,1,\dots,1)^\intercal $ is a left-eigenvector of $M$ with eigenvalue 1. Now, by the Perron-Frobenious Theorem, 
 there are no other positive left-eigenvectors, and all other eigenvalues of $M$   have norm strictly less than 1. Then $\lim_{n\to \infty}M^n=\bm{v}\bm{w}^\intercal$, where $\bm{v}$ is a (properly) normalized Perron right eigenvector, whose precise form is not important. 
 Lastly, because $\bm{a}_0$ is orthogonal to $\bm{w}$, we obtain $\lim_{n\to \infty} {\boldsymbol a}_n =\bm{v}\bm{w}^\intercal\bm{a}_0={\boldsymbol 0}$.\end{proof} 

We have shown that the ITAC is right-invariant under rotation by limits of subsequences of $U(S_n)$. We now show that it also right-invariant by any other unitary which is generated by these limits.

\begin{lemma}
\label{lemm:Iisaclosedgroup}
The set of unitaries which leave all the ITACs invariant, $$\mathbb{G}=\{U\in \SU(d)\mid \forall k\in\mathbb{N}\colon \mathcal{N}_\infty \circ \mathcal{U}= \mathcal{N}_\infty\}$$
is a closed subgroup of $\SU(d)$.
\end{lemma}
\begin{proof}
 It is easy to see that if $U,V\in \mathbb{G}$, then $UV,V^{\dagger}\in \mathbb{G}$, making it a subgroup. Because $U\mapsto \mathcal{U}$ is continuous, the set of unitaries $U$ such that $\mathcal{N}_\infty \circ \mathcal{U}= \mathcal{N}_\infty$ for a fixed $k$ is closed. Taking the intersection over all $k$ gives $\mathbb{G}$, which then must be closed as well.
\end{proof}

 From Lemmas  \ref{lem:subsequentiallimitsinvariant} and \ref{lemm:Iisaclosedgroup}, to show Eq.~\eqref{eq:setAequivance}, it would be enough to show that there exists a pair of limits of subsequences of ${U}(S_n)$ that generate a dense subgroup of $\SU(d)$.  This is indeed the case, and in fact they will be precisely the unitaries $A_0$ and $A_1$ that were used to generate the evolution. Why would the unitaries  $A_0,A_1$ generate all possible other unitaries?  This  is well known in the quantum computation community \cite{Lloyd1995}: one can build a universal quantum computer with almost any two gates sampled at random from $\SU(d)$.
 \begin{lemma}[Almost any pair of quantum logic gates is universal \cite{Lloyd1995}]
 \label{lemm:univcompprob1}
 If $A_0,A_1$ are sampled from the Haar measure of $\SU(d)$, then
     $$\Pr[\langle A_0,A_1\rangle\text{ is dense in }\SU(d)]=1.$$
     where $\langle A_0,A_1\rangle$ is the group generated by $ A_0$ and $A_1$.
 \end{lemma}

From Lemma  \ref{lemm:univcompprob1}, it follows that, in order to find a pair of limits of subsequences of ${U}(S_n)$ that generate a dense subgroup of $\SU(d)$, it is enough to show that, with unit probability, $A_0$ and $A_1$ are limits of subsequences of $U(S_n)$. This will follow from a well known result:

{\noindent\bf Poincaré Recurrence Theorem (PRT).}
{\it Let $X$ be a metric space with a Borel probability measure, and let $\Phi:X\to X$ be a continuous measure preserving transformation. With unit probability,  $x\in X$ is the limit of a subsequence of $(\Phi^n(x))_{n\in \mathbb{N}}$. 
 } (c.f. \cite[2.2.3]{Einsiedler2010}).
The physical intuition is that, under the dynamics induced by $\Phi$, except for a zero-measure set, all points are {\it recurrent}, meaning that they come back arbitrarily close to themselves, infinitely often. 

 \begin{lemma}[Recurrence of the Fibonacci dynamics] \label{lemma:recurranceA0A1} If $A_0,A_1$ are sampled uniformly from $\SU(d)$
 $$\Pr[A_0\textrm{ and } A_1\textrm{ are limits of subsequences of }U(S_n)]=1.$$ 
 \end{lemma}

 \begin{proof}

 Consider the map $\Phi\colon\SU(d)^2\to \SU(d)^2$ given by $\Phi(V,W)= (W,VW^m)$, which satisfies
 $\Phi^{n}(A_1,A_0)=(U(S_{n}),U(S_{n+1}))$
 and is Haar-measure preserving \cite[59.B]{Halmos1950}. By the PRT, the set of points $(A_1,A_0)\in \SU(d)^2$ that are subsequential limits of the sequence $(U(S_{n}),U(S_{{n}+1}))_{n\in\mathbb{N}}$ 
   has full Haar measure. 
 \end{proof}

Finally, we can prove our main result, putting all four preceding lemmas together.
\begin{proof}[Proof of Theorem 1]
On the one hand, Lemma \ref{lem:subsequentiallimitsinvariant} says that any limit of a subsequence of $U(S_n)$ leaves the ITAC invariant, i.e., is in the group $\mathbb{G}$ defined in Lemma \ref{lemm:Iisaclosedgroup}. Thus, from Lemma \ref{lemma:recurranceA0A1}, we have  $$\Pr[\{A_0,A_1\}\subseteq  \mathbb{G}]=1.$$ On the other hand, Lemma \ref{lemm:univcompprob1}  guarantees that $$\Pr[\langle A_0,A_1\rangle\text{ is dense in }\SU(d)]=1.$$ By joining both we obtain $$\Pr[\langle \mathbb{G}\rangle\text{ is dense in }\SU(d)]=1.$$ Lemma \ref{lemm:Iisaclosedgroup} says that $\mathbb{G}$ is a closed group, so $\mathbb{G}=\langle\mathbb{G}\rangle$ can only be dense if $\mathbb{G}=\SU(d)$, and then we can conclude \begin{equation*}1=\Pr[\mathbb{G}=\SU(d)]=\Pr[\forall k\in\mathbb{N}\colon\, \,\forall{U\in \SU(d)}\colon \mathcal{N}_\infty \circ \mathcal{U}= \mathcal{N}_\infty]=\text{Pr}[\forall{k\in \mathbb{N}} : \mathcal{N}_\infty=\mathcal{N}_\mathrm{Haar}].\qedhere\end{equation*}
\end{proof}

\section{Numerical methods}
\label{app:numerics}
In the main text, we present numerical simulations for the Fibonacci drive of order $m=1$, for which we compute the trace distance $\Delta^{(k)}(T)\coloneqq \tfrac{1}{2}\norm*{\rho_T^{\scriptscriptstyle(k)}-\rho_{\mathrm{Haar}}^{\scriptscriptstyle(k)}}_1$. In this section, we explain our numerical methods in detail.

The matrix $\rho_{\mathrm{Haar}}^{\scriptscriptstyle(k)}$ can be easily obtained directly from Eq.~\eqref{eq:haarensemble}. The calculation of $\rho_T^{\scriptscriptstyle(k)}$, however, can be more computationally expensive, especially for large times $T$. The recursive relations given by Eq.~\eqref{eq:recursiveForumla} allow efficient computation of $\rho_T^{{\scriptscriptstyle(k)}}$ at the times  $T=S_{n}$ given by Eq.~\eqref{eq:SnRecurrence}. Under the following scheme, computing $\rho_{T=S_{n}}^{\scriptscriptstyle (k)}={\mathcal{N}}^{\scriptscriptstyle(k)}_{S_n}[\dyad{\psi}^{\smash{\otimes k}}]$ requires only $n$ operations. To achieve this we compute a matrix representation of the channel $\mathcal{N}^{\scriptscriptstyle(k)}_{S_n}$, which we denote by $\text{mat}(\mathcal{N}^{\scriptscriptstyle(k)}_{S_n})$, which then acts on the vectorization of the initial state $\text{vec}(\dyad{\psi}^{\smash{\otimes k}})$. Explicitly, one represents the unitary channel $\mathcal{U}$ by $\text{mat}(\mathcal{U})=U^{\smash{\otimes k}}\otimes U^{\smash{*\otimes k}}$, and the channel composition by matrix multiplication $\text{mat}(\mathcal{N}\circ \mathcal{U}) = \text{mat}(\mathcal{N})\text{mat}({\mathcal{U}})$. Equation~\eqref{eq:recursiveForumla} then becomes  
    \begin{align*}
\text{mat}({\mathcal{N}}^{\scriptscriptstyle(k)}_{S_0})&=\text{mat}({\mathcal{N}}^{\scriptscriptstyle(k)}_{S_1})=\mathds{1}^{\otimes 2k} \nonumber \\
\text{mat}(\mathcal{N}^{\scriptscriptstyle(k)}_{S_{n+1}})&=\sum_{y=0}^{m}\frac{S_{n-\delta_{ym}}}{S_{n+1}}\,\text{mat}(\mathcal{N}_{S_{n-\delta_{ym}}})\big(U(S_{n})^{\otimes k}\otimes U(S_{n})^{*\otimes k}\big)^{ y}.\nonumber
    \end{align*}
    
One can apply the matrix $\text{mat}(\mathcal{N}^{\scriptscriptstyle(k)}_{S_n})$ of size $d^{2k}\times d^{2k}$ like an operator to the vectorized state $\text{vec}(\dyad{\psi}^{\smash{\otimes k}})$ to find $\text{vec}(\rho^{{\scriptscriptstyle (k)}}_{{S_n}})=\text{mat}(\mathcal{N}^{\scriptscriptstyle(k)}_{S_n})\text{vec}(\dyad{\psi}^{\smash{\otimes k}})$. This is efficient to compute in time at the cost of performing operations on $d^{2k}$-dimensional complex matrices. This is the method utilized to obtain the results shown in Figs.~3b and 3c of the main text. Importantly, the quantity $\Delta^{(k)}$ may become smaller than the floating point precision, so one has to increase the decimal precision of the matrices to obtain accurate results. A convenient way to quantify the numerical error is to compute the quantity $\varepsilon_n=\norm*{\mathds{1}-U(S_n)U(S_n)^\dagger}_\infty$, where $\norm*{O}_\infty$ is the operator norm, equal to the maximum eigenvalue of $O$. For all our numerical analyses, we use a numerical precision that guarantees that $\varepsilon_n	\ll \Delta^{(k)}(S_n)$. 

For the results shown in Fig.~3d in the main text (many-body case), the  method described above is not efficient, as $d=2^{\smash{L}}$ becomes too large. Nevertheless, we use the recursive structure to parallelize the computation in time. We split the time window into $n$ chunks, $1<T_1<T_2<\dots< T_n$.  We write $T_j=\sum_{i=1}^{\smash{J}} F_{c_i}$ as a sum of distinct Fibonacci numbers $F_{c}=S_{c+1}\quad (m=1)$, called the Zeckendorf representation, where $c_{i+1}> c_i+1\geq 3$. Then it can be shown that $U(T_j)=U(F_{c_1})U(F_{c_2})\cdots U(F_{c_J})$, where $U(F_{c_i})$ can be computed efficiently using Eq.~\eqref{eq:recursiveForumlaoperator}, with $m=1$. Using this formula, we may efficiently compute $\ket{\psi(T_j)}=U(T_j)\ket{\psi(0)}$ for exponentially large $T$. Then, in parallel, we apply the gates $A_0,A_1$ in the Fibonacci sequence within each time window $[T_j,T_{j+1}]$ and average the $k$th moments, to finally average over all the results for each time window.

\section{Further numerical analysis of the single-qubit Fibonacci drive}
\label{app:qubit}
\begin{figure}[h]
    \centering
    \includegraphics[width=0.20\columnwidth]{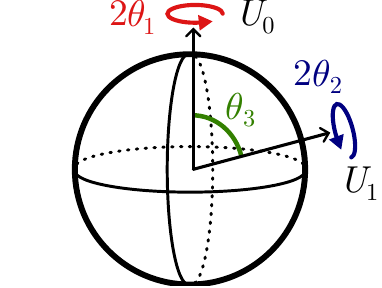}
    \caption{Single-qubit rotations $U_0$ and $U_1$.}
    \label{fig:S1}
\end{figure}

In this section, we analyze in detail the behavior of a single qubit under the evolution given by the Fibonacci drive of order $m=1$, beyond what we present in the main text. In the main text, we considered a pair of rotations, one around the $Z$-axis and the other around the $X$-axis. Here we keep the rotation around the $Z$-axis by an angle $2\theta_1$, $U_0=\exp\left(-i\theta_1 Z\right)$, but we generalize the second unitary to be a rotation by an angle $2\theta_2$ around an axis that forms an angle $\theta_3$ with the $Z$ axis, $U_1=\exp\left[-i\theta_2 \left(\cos\theta_3 Z+\sin\theta_3 X\right) \right]$, as shown in Fig.~\ref{fig:S1}. This parametrization captures all possible pair of single-qubit unitaries up to a global rotation.

As in the main text, we apply the Fibonacci drive of order $1$ using this pair of unitaries to a random initial state (the results are independent of the initial state). We consider the trace distance $\Delta^{(k)}(t)$, and fit a power law $\Delta^{(k)}(t)\sim e^{-\gamma t}$. We provide more details on the accuracy of this fit in the next subsection. At $\theta_3=\pi/2$ we recover the model presented in the main text, where the colormap for $\gamma$ is depicted in Fig.~\ref{fig:S2}a1. The features observed for the parameter choice $\theta_3=\pi/2$ roughly remain the same when we slightly move away from to $\theta_3=0.45\pi$ (Fig.~\ref{fig:S2}a2). However, there are significant changes as we vary $\theta_3$ further to $0.4\pi$.

In Fig.~\ref{fig:S2}a3 we observe that there exists a finite region  where $\gamma=1$ (white region with star marker). Points inside this region follow a power law decay $\Delta\sim 1/t$ which is plotted in Fig.~\ref{fig:S2}c with a dark red line. An almost perfectly linear decay for $\Delta$ in time is intriguing, as it beats the decay $\gamma=1/2$ that one would obtain from random sampling. We find that this occurs only when $k=2$. Repeating the calculations for $k=3$ (Fig.~\ref{fig:S2}a4), reveals a maximum value of $\max(\gamma) \sim 0.6$ for $\theta_3=0.4\pi$, and, interestingly, the region where we found $\gamma_{k=2} \sim 1$  turns out to have vanishingly small decay $\gamma_{k=3} \sim 0.06$, which is plotted with a light color in Fig.~\ref{fig:S2}c, and marked with a diamond in Fig.~\ref{fig:S2}a4. For angles inside this region, the evolution quickly approximates a $k=2$ design, but consequently fail to quickly approximate a $k=3$ design. Providing an analytical understanding of this effect is an interesting problem which we leave open.
\subsection{Convergence properties of single-qubit simulations}

\begin{figure*}
\centering
\includegraphics[width=0.99\textwidth]{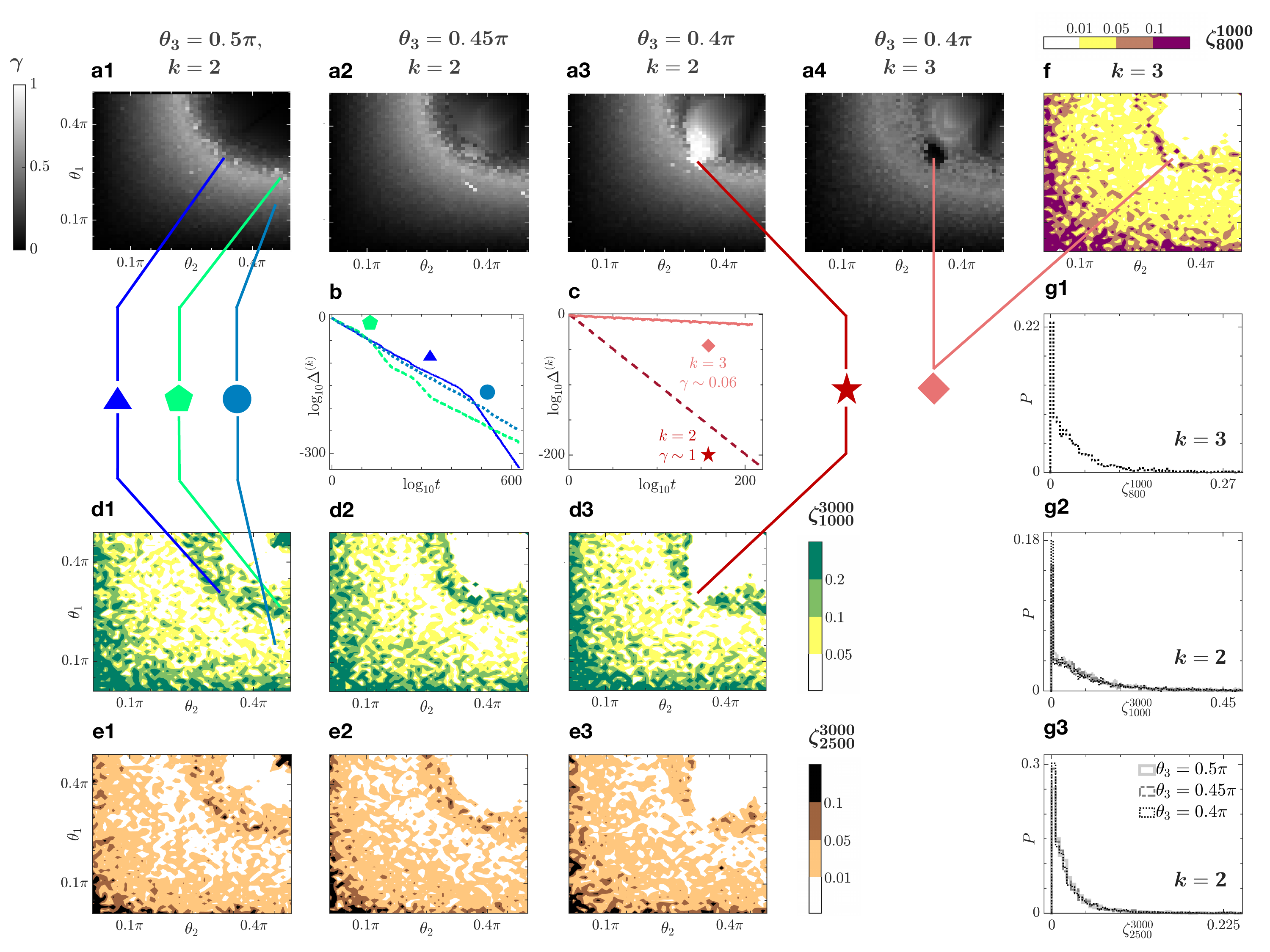}\hfill
\caption{\label{fig:S2} \textbf{a} Power-law exponent $\gamma$ map for $k=2$ and  a1 $\theta_3=0.5\pi$, a2 $\theta_3=0.45\pi$, a3 $\theta_3=0.4\pi$ time-averaged up to the Fibonacci time $F_{3000}=4.106 \times 10^{626}$, and a4 $k=3$, $\theta_3=0.4\pi$ time-averaged up to the Fibonacci time $F_{1000}=4.35\times 10^{208}$. \textbf{b} Finite-time trace distance $\Delta^{k=2}$ for three characteristic points chosen on a1, shown with a triangle, pentagon and circle. \textbf{c} Finite-time trace distances, $\Delta^{k=2}$ for the point $(\theta_1=0.30\pi,\theta_2=0.33\pi,\theta_3=0.4\pi)$ with maximum $\gamma\sim1$, marked on a3 with a star and $\Delta^{k=3}$ for the same point, marked on a4 with a diamond.
\textbf{d} $\zeta^{3000}_{1000}$ for d1 $\theta_3=0.5\pi$, d2 $\theta_3=0.45\pi$, d3 $\theta_3=0.4\pi$, and $k=2$. \textbf{e} $\zeta^{3000}_{2500}$ for e1 $\theta_3=0.5\pi$, e2 $\theta_3=0.45\pi$, e3 $\theta_3=0.4\pi$,  and $k=2$. \textbf{f} $\zeta^{1000}_{800}$ for $\theta_3=0.4\pi$ and $k=3$. \textbf{g} Histograms of g1 $\zeta^{1000}_{800}$, $k=3$, 
 g2 $\zeta^{1000}_{3000}$, $k=2$ and g3 $\zeta^{2500}_{3000}$, $k=2$.}
\end{figure*}

We now analyze the convergence properties of the power-law decay of $\Delta^{(k)}$ observed in single-qubit simulations of the main text and the preceding section. When we perform a numeral fit, we only compute the evolution up to a finite time $\tau$. The obtained exponent $\gamma$ depends on $\tau$, i.e., $\Delta^{(k)} \sim t^{-\gamma(\tau)}.$ Of course, if the infinite-time behavior is truly a power-law, then $\gamma(\tau)$ should converge to a fixed number for sufficiently large values of $\tau$. We can learn if the numerical value $\gamma(\tau)$ is well converged by analyzing its dependence on $\tau$. For this purpose, we consider the quantity
\begin{eqnarray}
    \zeta_{n}^{m} \coloneqq\frac{|\gamma(\tau=F_{n})
   -\gamma(\tau=F_{m})|}{\gamma(\tau=F_m)},
\end{eqnarray}
where $m>n$ and $F_m$ represents the $m$'th Fibonacci number. A value of $ \zeta_{n}^{m} =0$ means that $\gamma(F_{n})=\gamma(F_{m})$, and generically the value $\zeta_{n}^{m}$ may be read as the relative change from $\gamma(F_{m})$ to $\gamma(F_{n})$.

We set a threshold for convergence to be $\zeta_n^m <0.1$, meaning that the relative change in $\gamma(\tau)$ is below 10\% from $F_n$ to $F_m$.
Our data runs up to $F_{3000}=4.106 \times 10^{626}$, so we study two different values for $\zeta_{n}^{m}$: $\zeta_{2500}^{3000}$  and $\zeta_{1000}^{3000}$. Let us also note that we find no significant dependence of the following results if we also change the initial time used for the fitting. 

We plot $\zeta_{1000}^{3000}$ for three different values of the parameter $\theta_3$ in Fig.~\ref{fig:S2}d. Most of the points on these convergence maps feature a value of $\zeta_{1000}^{3000} < 0.1$ (yellow and white regions). The histogram of $\zeta_{1000}^{3000}$ over the $\theta_1$ and $\theta_2$ plane given in Fig.~\ref{fig:S2}g2 shows a distribution close to Poisson,  peaked at $\zeta \in [0,0.005]$, with $63.6$\%, $67.1$\%, $67.4$\% of the values under the $\zeta_{1000}^{3000} < 0.1$ threshold for $\theta_3=0.5\pi$, $\theta_3=0.45\pi$ and $\theta_3=0.4\pi$, respectively. For $\zeta_{2500}^{3000}$, this number increases substantially. The histogram (Fig.~\ref{fig:S2}g3) shows that a total of $96.3$\%, $96.4$\% and $96.9$\% of points satisfy $\zeta_{2500}^{3000} < 0.1$, for $\theta_3=0.5\pi$, $\theta_3=0.45\pi$ and $\theta_3=0.4\pi$, respectively. Most of the darkest points in Figs.~\ref{fig:S2}e, where $\zeta_{2500}^{3000}>0.1$, reside at the edges of the planes, i.e.,~$\theta_1 \sim 0$ or $\theta_2 \sim 0$. This is expected, as $\gamma$ is very small close to these values, so $\zeta_n^m$ grows larger.

We observe important differences on the convergence maps for $\theta_3=0.5\pi$ and away from it. As the histograms display, the peak of the distribution  becomes sharper as we move away from $\theta_3=0.5\pi$ to $\theta_3=0.4\pi$. Therefore, the exponent maps for $\theta_3 \neq 0.5\pi$ have more points that exhibit stable and convergent power-law decay. The most significant region of the convergence maps is the region where the quarter-ring feature lies, where we observe non-generic decay behaviors. To analyze in detail, we choose three specific points on the map for $\theta_3=0.5\pi$, corresponding to Figs.~\ref{fig:S2}a1, \ref{fig:S2}d1, \ref{fig:S2}e1. The first point is depicted with a triangle at $\theta_1=0.3\pi$ and $\theta_2=0.33\pi$. This point corresponds to the darkest region in both convergence maps (Figs.~\ref{fig:S2}d1, \ref{fig:S2}e1), having the highest value of $\zeta$. Fig.~\ref{fig:S2}b shows the  point marked with a triangle, exhibiting a sudden change in the decay exponent at $t\approx 10^{450}$. The second point is depicted with a pentagon at $\theta_1=0.26\pi$ and $\theta_2=0.47\pi$ which corresponds to the intermediate region in both convergence maps. Specifically this region has $0.1 \leq \zeta_{1000}^{3000} < 0.2$ and $0.05 \leq \zeta_{1000}^{3000} < 0.1$. As Fig.~\ref{fig:S2}b shows, the decay exhibits a staircase behavior in the first half of the time evolution before stabilizing. The third point at $\theta_1=0.15\pi$ and $\theta_2=0.46\pi$ lies on a white region in both convergence maps, meaning that the power-law decay at this point converges well. Indeed, the plot marked with a circle  in Fig.~\ref{fig:S2}b visually confirms this observation. We note that the behavior of the points depicted with the triangle and the pentagon is rare, and the majority of points have a well-converged power law similar to the point marked with a circle.

For completeness, we provide a convergence map of $\zeta_{800}^{1000}$ for $\theta_3=0.4\pi$ and $k=3$ in Fig.~\ref{fig:S2}f and its histogram in Fig.~\ref{fig:S2}g1. Increasing $k$ does not eliminate the presence of the darkest region on the map, where $\zeta_{800}^{1000} \geq 0.1$. The number of points that above our threshold  $\zeta_{800}^{1000} \geq 0.1$ is $11.5$\%. We believe, however, that this number would substantially decrease by performing longer-time simulations.

\section{Deep Thermalization in Completely ergodic Dynamics}
\label{app:deep_thermalization}

In this section, we explain how quantum (deep) thermalization follows directly from CHSE. We also perform a numerical analysis to study the speed at which this phenomenon ensues.

Quantum thermalization is explained by considering a bipartite system, $\mathcal{H}=\mathcal{H}_A\otimes\mathcal{H}_B$, with $d_A=\text{dim}(\mathcal{H}_A)$ much smaller than $d_B=\text{dim}(\mathcal{H}_B)$. If we drive the entire system $\mathcal{H}$ under completely-ergodic time-dependent dynamics, and sample any evolved state  at a late time $t_\text{late}$, $\ket{\psi(t_\text{late})}\in \mathcal{H}$ is indistinguishable from a Haar-random vector. This is just the definition of CHSE. Now, by a measure-concentration argument, it can be shown that the reduced density matrix $\rho_B=\tr_B(\dyad{\psi}))$ of a Haar random vector $\ket{\psi}$ equals the maximally mixed state $\mathds{1}_A/d_A$ in the thermodynamic limit $d_B/d_A \to \infty$ \cite{Popescu2006}. Furthermore, a stronger physical process called deep thermalization \cite{Ippoliti2022} can be proven by a similar argument, as we show next.

To define deep thermalization, one considers the measurement of a global wavefunction $\ket{\psi}\in \mathcal{H}$ over an orthonormal basis $\ket{\beta}$ for the system $\mathcal{H}_B$, which yields a measurement outcome $\beta\in\{0,1,\dots, d_B-1\}$ with probability $p_\beta=\sum_{\alpha=0}^{\smash{d_A-1}}\abs*{\braket{\alpha,\beta}{\psi}}^{\smash{2}}$, where $\ket{\alpha}$ forms an orthonormal basis for $\mathcal{H}_A$. If the measurement outcome is $\beta$, the state on $\mathcal{H}_A$ gets updated to
\begin{eqnarray}
    \ket{\psi_\beta} = \frac{1}{\sqrt{p_\beta}}\left( \mathds{1}_A \otimes \bra{\beta}\right) \ket{\psi}.
\end{eqnarray}
The ensemble consisting of all measurement outcomes is called the projected ensemble \cite{Ho2022,Cotler2023},
\begin{eqnarray}
    \mathcal{E} \coloneqq  \{ (p_\beta,\dyad{\psi_{\beta}}) \}_\beta.
\end{eqnarray} The first moment of $\mathcal{E}$ is the reduced density matrix $\rho_A$, and thermalization (to infinite temperature) is the statement that $\rho_A(t)=\mathds{1}/d_A$,
at late times $t$.  Deep thermalization generalizes this idea to higher moments beyond the average, requiring that at late times $t$, the $k$th moment
\begin{equation}
\label{kthmomentthermalization}
    \rho^{(k)}_{\mathcal{E}}(t)=\sum_{\beta=0}^{d_B} p_\beta \dyad{\psi(t)_{\beta}}^{\otimes k}
\end{equation} converges to the Haar moment $\rho^{\scriptscriptstyle{(k)}}_{\mathrm{Haar}}$ [Eq.~\eqref{eq:haarensemble}] in $\mathcal{H}_A$, in the thermodynamic limit. 

Under CHSE dynamics, deep thermalization is guaranteed, which can be rigorously shown using Theorem 2 of Ref.~\cite{Cotler2023}. Specifically, we have the following.
\begin{theorem}
\label{thm:thmdeepthermalization}
    Let $H(t)$ be a time-dependent Hamiltonian on $\mathcal{H}_A\otimes \mathcal{H}_B$ which satisfies CHSE. For any initial state, $\ket{\psi}$, $k\in\mathbb{N}$ and  $\varepsilon,\delta>0$, there is a time $T>0$ such that, if $t$ is sampled uniformly at random from $[0,T]$, then \begin{equation}
        \label{eq:thmdeeptherm}
    \mathrm{Pr}\big[\tfrac{1}{2}\norm*{\rho^{(k)}_\mathcal{E}(t) - \rho_{\mathrm{Haar}}^{(k)}}_1<\varepsilon\big]>1-\delta,\end{equation} as long as 
    \begin{align}
    \label{eq:conds1}
\log(d_B)&=\Omega\!\left(k \log(d_A)+\log(\tfrac{1}{\varepsilon\delta})\right)&
&&\text{and}&&\log(d_A)&=\Omega\!\left( \log\log(d_B)+\log(k)+\log\log(\tfrac{1}{\varepsilon\delta})\right),
    \end{align}
    where $\Omega(\,\cdot\,)$ denotes a lower bound up to subleading terms and a multiplicative constant factor (see Ref. \cite{Cotler2023} for details).
\end{theorem}
\begin{proof}
    Let $k'$ sufficiently large and $\varepsilon'>0$ sufficiently small, such that they satisfy the requirements in Theorem 2 of Ref. \cite{Cotler2023}, i.e., 
       \begin{align}
         \label{eq:conds2}
k'&=\Omega\!\left(k \left(\log(d_B)+\log(\tfrac{1}{\varepsilon\delta})\right)\right)&
&&\text{and}&&-\log(\varepsilon')&=\Omega\!\left( k\log(d_B)\left(\log(d_B)+\log(\tfrac{1}{\varepsilon\delta})\right)\right).
    \end{align}
CHSE implies that there is some time $T$ such that the uniform ensemble $\{\ket{\psi(t)}\,|\,t\in[0,T]\}$ forms an $\varepsilon'$-approximate $k'$-design, meaning that $\tfrac{1}{2}\norm*{\rho_T^{(k')}-\rho_{\mathrm{Haar}}^{(k')}}_1<\varepsilon'$.  Theorem 2 of Ref. \cite{Cotler2023} then guarantees Eq. \eqref{eq:thmdeeptherm}.
\end{proof}

Theorem \ref{thm:thmdeepthermalization}  guarantees deep thermalization under CHSE dynamics, in the thermodynamic limit, where the dimension of both the subsystem $\mathcal{H}_A$ and its complement $\mathcal{H}_B$ are large, i.e., $d_A,d_B\to \infty$, but $\mathcal{H}_B$ is exponentially larger, i.e., $d_A\sim \log(d_B)$. However, this result does not give us information about the speed at which the system deep thermalizes. To study the rate of deep thermalization, we perform a numerical analysis.

\subsection{Numerical analysis of deep thermalization in the Fibonacci drive}
\begin{figure}
\centering
{\includegraphics[width=0.99\textwidth]{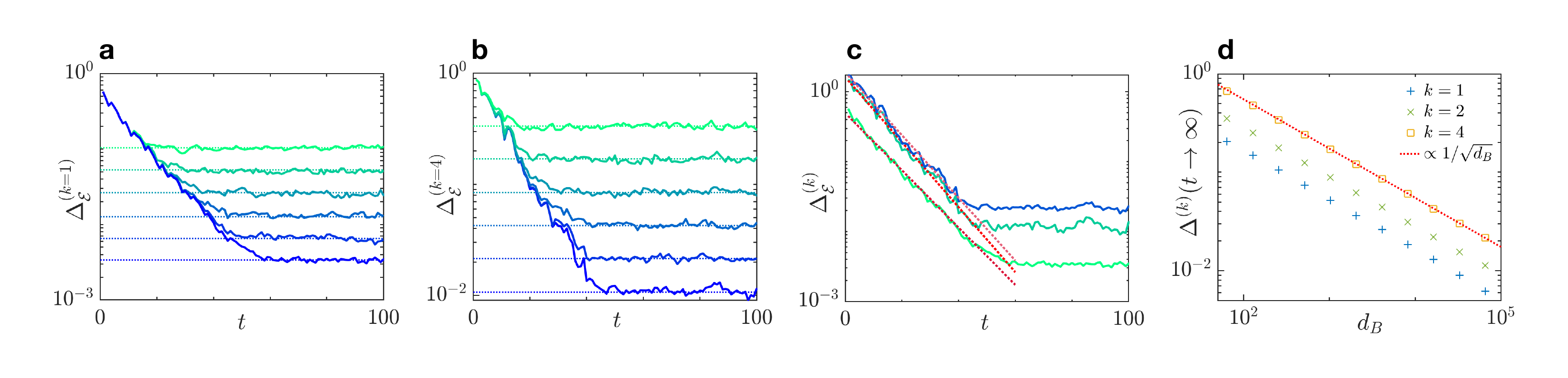}}\hfill
\caption{\label{fig:S3} {\textbf{a}} The trace distance $\Delta_\mathcal{E}^{\scriptscriptstyle{(k)}}$ probes  thermalization ($k=1$) after being averaged over $10$ initial random product states to decrease the fluctuations. {\textbf{b}}  $\Delta_\mathcal{E}^{\scriptscriptstyle{(k)}}$ probes deep thermalization ($k=4$) for only one initial random product state. In both a and b, the size of the subsystem $A$ is $N_A=2$ and the system sizes are $N=8,10,12,14,16,18$ with colors light cyan to dark blue, respectively. The horizontal dashed lines show $\Delta_\mathcal{E}^{\scriptscriptstyle (k)}$ for Haar-random states.
{\textbf{c}} Comparison of $\Delta_\mathcal{E}^{\scriptscriptstyle{(k)}}$ for different $k=1,2,4$ from light cyan to dark blue, respectively at $N=18$. The dotted red lines are the exponential fits to the decays $\Delta_\mathcal{E}^{\scriptscriptstyle{(k)}} \propto e^{-\lambda_{k}t}$ giving $\lambda_{1}\sim 0.09$, $\lambda_{2}\sim 0.11$ and $\lambda_{4}\sim 0.1$. {\textbf{d}}  The saturation value of $\Delta^{(k)}_\mathcal{E}$ decreases as a power-law in the dimension of $\mathcal{H}_B$, $\propto 1/\sqrt{d_B}$.}
\end{figure}
To study the rate of deep thermalization, we consider the same many-body system presented Fig.~3d of the main text: a spin-$1/2$ chain of length $L$, driven by the Fibonacci drive of order $m=1$, with $A_0=e^{-i H_0 \tau}$ and $A_1=e^{-i H_1 \tau}$, with $\tau=1$ and
\begin{align*}
    H_0=\sum_{j=1}^L X_j+\sum_{j=2}^L X_{j-1}X_j+X_L/10,&&
    H_1=\sum_{j=1}^L Z_j+\sum_{j=2}^L Z_{j-1}Z_j+Z_L/10.
\end{align*} We select the subsystem $\mathcal{H}_A$ to consist of the first $N_A=2$ qubits, so that $d_A=2^2$, and $\mathcal{H}_B$ to consist of the remaining $N_B=L-N_A$ qubits, so $d_B=2^{L-2}$. Finally, we construct $\rho^{\scriptscriptstyle (k)}_\mathcal{E}(t)$ via Eq.~\eqref{kthmomentthermalization} with initial state $\ket{\psi(0)}=\ket{\phi}^{\otimes L}$, where $\ket{\phi}$ is a random qubit state. We measure the distance to the Haar ensemble with the trace distance
\begin{eqnarray}
    \Delta^{(k)}_\mathcal{E}  \coloneqq  \frac{1}{2}\norm{\rho^{(k)}_\mathcal{E} - \rho_{\mathrm{Haar}}^{(k)}}_1.
\end{eqnarray}

Figs.~\ref{fig:S3}a and \ref{fig:S3}b show that $\Delta_\mathcal{E}^{\scriptscriptstyle(k)}$ exponentially decreases in time for $k=1$ and $k=4$ until it saturates to a finite-size plateau whose value is found to be $\propto 1/\sqrt{d_B}$ in Fig.~\ref{fig:S3}d. This is in contrast to the power-law decay found in chaotic, but time-independent dynamics \cite{Cotler2023}. For all the values of $k$ we present, between $1$ and $4$, we find $\Delta^{\scriptscriptstyle(k)}_{\mathcal{E}}\sim e^{-0.1t}$, as displayed in Fig.~\ref{fig:S3}c. The horizontal dashed lines in Figs.~\ref{fig:S3}a and~\ref{fig:S3}b show the average value of $\Delta_\mathcal{E}^{\scriptscriptstyle (k)}$ for Haar-random states, which corresponds to the value of the of the long-time plateau of $\Delta_\mathcal{E}^{\scriptscriptstyle(k)}(t)$, for initial product states. This is a consequence of the CHSE: at late times, any state is statistically indistinguishable from a Haar-random vector.

As a final remark, due to the relation $\Delta^{\scriptscriptstyle(k)}_{\mathcal{E}} \geq \textrm{Tr}[ {O}(\rho^{\scriptscriptstyle(k)}_{\mathcal{E}}-\rho^{\scriptscriptstyle(k)}_{\mathrm{Haar}}) ]$ for a normalized operator $\norm*{{O}}_{\infty}= 1$, our results imply exponentially fast thermalization of local observables.

\bibliography{references}